# Inter-defect interactions, oxygen-vacancy distribution, and oxidation in acceptor-doped $ABO_3$ perovskites

L.P. Putilov*, M.Z. Uritsky, V.I. Tsidilkovski*

*Institute of High-Temperature Electrochemistry, 20 Akademicheskaya St., Ekaterinburg 620990, Russia*

*corresponding authors:
L.P. Putilov, lev.putilov@gmail.com;
V.I. Tsidilkovski, vtsidilkovski@gmail.com

## Abstract

The effects of inter-defect interaction and impurity disorder on defect thermodynamics, local ion coordination, and oxidation in acceptor-doped wide-gap $AB_{1-x}R_xO_{3-\delta}$ perovskites are explored using the developed statistical theory and Monte Carlo simulations. The results demonstrate that under realistic energy parameters the interaction between oxygen vacancies and impurities generally has a greater impact on the studied properties than inter-vacancy correlations. The influence of inter-vacancy interaction significantly depends on dopant content $x$: inter-site vacancy repulsion becomes noticeable at sufficiently high $x$, whereas on-site correlations can be pronounced within a narrow doping range at moderate $x$ values. It is found that a non-uniform impurity allocation, which can result from a sample preparation procedure, considerably affects oxygen-vacancy distribution, and has a weaker effect on short-range order and oxidation. It is also shown that inter-defect interaction reduces the hole concentration, increases the oxidation enthalpy, and can result in their non-trivial dependence on $x$. The obtained results agree with experimental data on local ion coordination and help to explain the behavior of hole conductivity in the considered perovskites. The findings of this study contribute to understanding the fundamental properties of acceptor-doped oxides, facilitating the development of new materials for clean energy applications.

## 1 Introduction

Acceptor-doped perovskites $AB_{1-x}R_xO_{3-\delta}$ have significant potential as promising materials for clean energy applications [1-4]. Acceptor doping of $ABO_3$ oxides leads to the formation of oxygen vacancies, which enables such materials to be employed as oxygen, hydrogen or ionic-electronic conductors in solid oxide electrochemical cells [5]. At the same time, acceptor impurities alter various material properties, in particular those important for applications. As has been established experimentally [6-11] and theoretically [12-20], interaction between charge carriers and acceptor impurities is one of the most important effects influencing the fundamental properties of the studied oxides. Numerous studies of these effects in proton-conducting perovskites have demonstrated that the interaction of protons and oxygen vacancies with impurity ions strongly affects hydration and transport phenomena (e.g., Refs. [7,17-22]). In our previous works, we showed that these interactions impact not only hydration, oxidation and transfer processes [17,19,23], but also the performance of electrochemical cells based on proton-conducting membranes [18]. The vacancy–impurity interaction itself has important implications for defect and transport phenomena, including oxidation, oxide-ion and hole conduction [23-27]. Previously, we addressed the effect of oxygen vacancy trapping on the oxidation of $AB_{1-x}R_xO_{3-\delta}$ perovskites assuming only one type of acceptor-bound vacancy states and uniform dopant distribution [18,23]. In particular, it was shown [18] that vacancy trapping noticeably affects the Fermi level position and hole concentration, which is essential for the transport properties of oxides and their applications in electrochemical devices.

The distribution of oxygen vacancies over lattice sites with different impurity surroundings should also be largely determined by the interactions between vacancies and impurities. The corresponding short-range order related to the local coordination of vacancies, in turn, affects the transport and thermodynamic properties of oxides. The effects of interactions and non-uniform dopant distribution on the local environments of ions in acceptor-doped perovskites were studied experimentally using nuclear magnetic resonance (NMR) technique [6,11,28,29]. Theoretically, the effect of frozen dopant configuration, corresponding to equilibrium at typical sintering temperatures (~ 1900 K), on the local structure of Y- and Sc-doped $BaZrO_3$ was explored using density functional theory (DFT) and Monte-Carlo simulations [30,31].

Despite numerous studies of inter-defect interactions in acceptor-doped perovskites, a number of fundamental issues related to the implications of these interactions, especially for heavily doped oxides, remain unresolved. Note that in applications, where the studied materials can serve as, e.g., proton-conducting electrolytes in electrochemical cells, oxides with a dopant content of up to ~ 20 % are typically employed. Recent studies reported perovskites that retain a single-phase structure at significantly higher impurity concentrations [10,32-34]. In such heavily doped oxides, the effect of non-ideality should be essential and can result in new consequences compared to lightly doped compounds. In particular, the interaction of vacancies with impurities and among each other should significantly affect their distribution over oxygen sites, local structure and oxidation. Furthermore, the effects of defect interactions depend on the distribution of acceptor impurities in oxides. However, the impact of this distribution on the properties under study, as far as we know, has not yet been investigated.

In this work, we explore the effects of inter-defect interaction and quenched impurity disorder on oxygen-vacancy distribution, local ion coordination, and oxidation in acceptor-doped wide-gap perovskites under dry oxidizing conditions. Using the developed statistical theory and Monte Carlo simulations, we investigate the following effects related to non-ideality of defect system. (1) The formation of

two- and three-particle complexes of acceptor-bound oxygen vacancies. (2) The impact of on-site Fermi-type correlations arising at high vacancy occupancy of oxygen sites surrounded by impurities. (3) The effects of inter-vacancy repulsion in nearest-neighbor positions. (4) The influence of non-uniform impurity distribution, which can result from a sample preparation procedure.

Our analysis of the above effects highlights the importance of inter-defect interaction for the defect and local structure of acceptor-doped $ABO_3$ oxides. We find that under realistic energy parameters the main effect on the studied properties is generally due to vacancy-impurity interactions. At the same time, short-range inter-vacancy correlations can provide a noticeable contribution to defect thermodynamics within a specific range of dopant concentrations. We show that inter-defect interaction considerably alters the oxide-gas equilibrium constant and oxidation enthalpy, and results in their strong dependence on dopant content. We also demonstrate that non-uniform impurity allocation can significantly influence the vacancy distribution, but has little effect on local coordination and oxidation behavior.

The results of the presented statistical thermodynamic analysis are in agreement with our Monte Carlo simulations. We also verify our findings on the defect and local structure of $AB_{1-x}R_xO_{3-\delta}$ perovskites using the NMR data for acceptor-doped $BaZrO_3$ [28,29]. In addition, our findings on oxidation help to explain the behavior of hole conductivity in Y- and In-doped $BaSnO_3$ [35,36].

# 2 Theory

## 2.1 The model

In this study, we consider wide-band-gap acceptor-doped perovskites $AB_{1-x}R_xO_{3-\delta}$ under dry oxidizing conditions. The concentration of holes in the oxide is assumed to be negligibly small compared to that of oxygen vacancies $c_V$ (see Section 3.5) and therefore $c_V \approx x/2 = \text{const}$. In the temperature range under consideration, the mobility of host and dopant cations is low and, accordingly, their distribution is supposed to be frozen.

Let us first consider a model that neglects inter-site interaction between vacancies and accounts only for on-site (Fermi-type) correlations, which are due to the prohibition of site occupation by more than one particle. The Hamiltonian for this model is

$$H_0 = \sum_{i=1}^{3N} E_{\mathrm{V}i} n_{\mathrm{V}i}, \tag{1}$$

where $n_{\mathrm{V}i}$ is the fermion occupation number of an oxygen site $i$ by a vacancy ($n_{\mathrm{V}i} = 1$ if site $i$ is vacant and $n_{\mathrm{V}i} = 0$ otherwise); $E_{\mathrm{V}i}$ is the vacancy energy at site $i$; $3N$ is the total number of oxygen sites in a crystal with $N$ unit cells.

Due to various possible dopant configurations in the region surrounding an oxygen vacancy V, the energy spectrum $\{E_{\mathrm{V}i}\}$ in $AB_{1-x}R_xO_{3-\delta}$ perovskites can be quite complex. In this work, however, to reveal the main effects of the considered interactions, we adopt a model with a reduced, three-level, energy spectrum. These levels – $E_{\mathrm{VR}}$, $E_{\mathrm{V2R}}$ and $E_{\mathrm{Vf}}$ – correspond to the energies of vacancies with different sets of the B-sublattice cations in the nearest environment: R-V-B, R-V-R, and B-V-B, respectively (see Fig. 1). The vacancies in these environments are denoted as $V_R$, $V_{2R}$, and $V_f$.

In what follows, we explore the properties of oxides as a function of the trapping energies of vacancies near one ($\Delta E_{VR}$) and two ($\Delta E_{V2R}$) impurities:

$$\Delta E_{\mathrm{VR}} = E_{\mathrm{Vf}} - E_{\mathrm{VR}},\ \Delta E_{\mathrm{V2R}} = E_{\mathrm{Vf}} - E_{\mathrm{V2R}}. \quad (2)$$

It is assumed that the effects of crystal and electronic structure reorganization due to the vacancy formation in various configurations are accounted for in $\Delta E_{VR}$ and $\Delta E_{V2R}$ [12-15].

To analyze the effects of inter-vacancy interaction, we consider a model with infinite repulsion between vacancies in nearest-neighboring sites (the simultaneous occupation of such sites by vacancies is prohibited). The corresponding Hamiltonian $H$ including the interaction term $H_{\mathrm{int}}$ can be specified as

$$H = H_0 + H_{\mathrm{int}} = H_0 + \lambda \sum_{\langle i,j \rangle} n_{\mathrm{Vi}} n_{\mathrm{Vj}}, \quad (3)$$

where $H_0$ is given by Eq. (1); $\lambda$ is the inter-vacancy repulsion energy, and the notation $\langle i,j \rangle$ implies a sum over all distinct pairs of the nearest-neighboring oxygen sites. The limit $\lambda \to \infty$ corresponds to the model with infinite repulsion, and $\lambda = 0$ is suitable for the model without inter-site vacancy interaction.

In what follows, for brevity, we refer to on-site Fermi-type correlations as weak, and correlations for the model with both on-site and inter-site vacancy repulsion as strong.

Since we consider a system with a quenched impurity disorder, finding the expectation value $\overline{L}$ of any observable $L$ implies double averaging [37,38]:

$$\overline{L} = \langle\langle L \rangle\rangle_{\omega}, \quad (4)$$

where $\langle\ \rangle$ is the usual Gibbs thermodynamic average and $\langle\ \rangle_{\omega}$ means averaging over the distribution $\omega$ of random variables characterizing disorder. In the case of a single-particle energy spectrum $\{E_{\mathrm{Vi}}\}$, the value $\overline{L}$ is given by

$$\langle\langle L \rangle\rangle_{\omega} = \int \prod_i [\omega(E_{\mathrm{Vi}}) dE_{\mathrm{Vi}}] \langle L(\{E_{\mathrm{Vi}}\}) \rangle, \quad (5)$$

where $\omega(E_{\mathrm{Vi}})$ is the distribution of energies corresponding to the impurity disorder, and $\langle L(\{E_{\mathrm{Vi}}\}) \rangle$ is the Gibbs average for a given set of energies $\{E_{\mathrm{Vi}}\}$.

For our model with three energy levels, the probability distribution $\omega$ takes the form

$$\omega(E_{\mathrm{Vi}}, x) = p_{\mathrm{R}} \delta(E_{\mathrm{Vi}} - E_{\mathrm{VR}}) + p_{\mathrm{2R}} \delta(E_{\mathrm{Vi}} - E_{\mathrm{V2R}}) + p_{\mathrm{f}} \delta(E_{\mathrm{Vi}} - E_{\mathrm{Vf}}), \quad (6)$$

where $p_R$, $p_{2R}$ and $p_f$ are the probabilities that a randomly chosen oxygen site has the corresponding local environment (R-O-B, R-O-R and B-O-B). Obviously, these probabilities depend on the dopant content $x$. In the case of the uniform distribution of impurities over B-sites, the probabilities are given by

$$p_{\mathrm{R}} = 2x(1-x),\ p_{\mathrm{2R}} = x^2,\ p_{\mathrm{f}} = (1-x)^2. \quad (7)$$

We explore the effects of inter-defect interactions on the oxide properties for both uniform and non-uniform dopant distributions. The implications of the non-uniform distribution are considered in Sections 3.4 and 3.5.

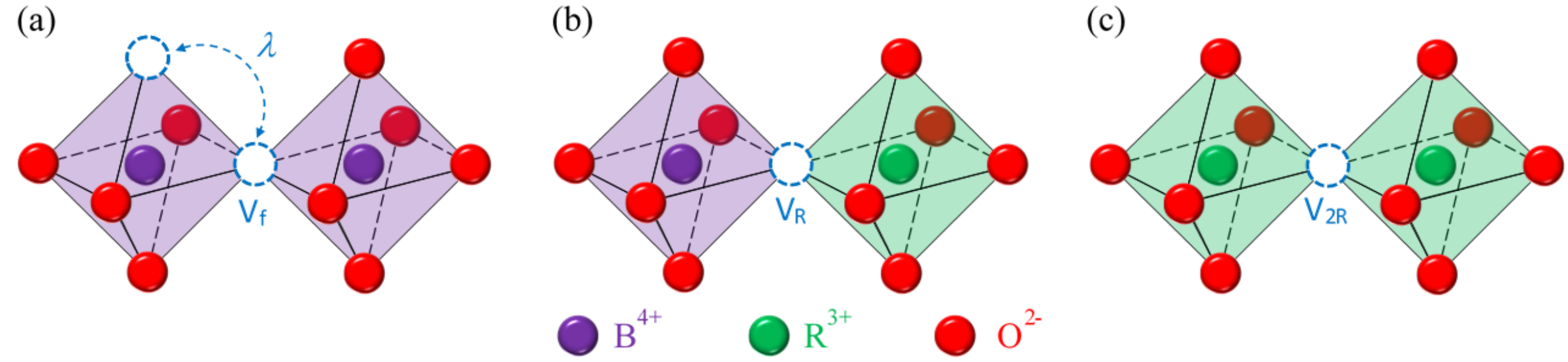

**Fig. 1.** Possible nearest environments of oxygen vacancies in acceptor-doped $AB^{4+}_{1-x}R^{3+}_{x}O_{3-\delta}$ perovskites: (a) B-V-B (b) B-V-R and (c) R-V-R. The inter-site vacancy interaction characterized by the repulsion energy $\lambda$ (see Eq. (3)) is schematically shown in (a).

### 2.2 Defect thermodynamics

To consider the equilibrium properties of the studied oxides, we employ: (1) analytical methods of statistical mechanics and (2) Monte Carlo simulations. The model neglecting inter-site vacancy repulsion is treated using both approaches, and the effect of this repulsion is studied only by Monte Carlo method (analytical treatment of the latter effect requires complex, poorly controlled approximations, and is beyond the scope of this work).

For the Hamiltonian $H_0$ (Eq. (1)) without inter-site vacancy interaction, the grand partition function $Z$ and equilibrium occupation probability $\langle n_{\mathrm{Vi}}\rangle$ are well-known (e.g., Ref. [39]):

$$Z = \prod_{i=1}^{3N} \mathrm{Tr}\left\{\exp\left[\frac{(\mu_{\mathrm{V}}-E_{\mathrm{Vi}})n_{\mathrm{Vi}}}{kT}\right]\right\} = \prod_{i=1}^{3N}\left[1+\exp\left(\frac{\mu_{\mathrm{V}}-E_{\mathrm{Vi}}}{kT}\right)\right], \quad (8)$$

$$\langle n_{\mathrm{Vi}}\rangle = \left[1+\exp\left(\frac{E_{\mathrm{Vi}}-\mu_{\mathrm{V}}}{kT}\right)\right]^{-1}, \quad (9)$$

where $\mu_{\mathrm{V}}$ is the chemical potential of oxygen vacancies.

Using Eq. (4) with the distribution $\omega(E_{\mathrm{Vi}}, x)$ (6), we obtain the averaged grand potential

$$\overline{\Omega} = \langle -kT\ln Z\rangle_{\omega} = -3NkT\sum_{m=1}^{3} p_{\mathrm{m}} \ln\left[1+\exp\left(\frac{\mu_{\mathrm{V}}-E_{\mathrm{Vm}}}{kT}\right)\right], \quad (10)$$

where $m$ denotes the type of local vacancy environment $m = \mathrm{V_f}$, $\mathrm{V_R}$ or $\mathrm{V_{2R}}$, corresponding to the configurations B-V-B, R-V-B and R-V-R.

The concentration of vacancies of type $m$ per formula unit is determined by the occupation probability $\langle n_{\mathrm{Vm}}\rangle$ (9) averaged over $\omega(E_{\mathrm{Vi}}, \mathrm{x})$ (6):

$$c_{\mathrm{Vm}} = 3p_{\mathrm{m}}\langle n_{\mathrm{Vm}}\rangle = 3p_{\mathrm{m}}\left[1+\exp\left(\frac{E_{\mathrm{Vm}}-\mu_{\mathrm{V}}}{kT}\right)\right]^{-1}. \quad (11)$$

The chemical potential $\mu_{\mathrm{V}}$ is determined by the self-consistency condition

$$c_{\mathrm{Vf}} + c_{\mathrm{VR}} + c_{\mathrm{V2R}} = c_{\mathrm{V}} = x/2, \quad (12)$$

which gives a cubic equation for the absolute activity $\exp(\mu_{\mathrm{V}}/kT)$.

To consider the effects of non-ideality in the vacancy system, we also use the activity coefficient $\gamma_{\mathrm{V}}$. We define $\gamma_{\mathrm{V}}$ in a standard way using $E_{\mathrm{Vf}}$ as the reference energy:

$$\mu_{\mathrm{V}} = E_{\mathrm{Vf}} + kT\ln\gamma_{V}\frac{c_{\mathrm{V}}}{3}. \quad (13)$$

The solution of Eqs. (11) and (12) yields the activity coefficient $\gamma_V$ (or $\mu_V$) and the concentrations $c_{Vm}$ that account for both trapping effects and Fermi-type correlations.

If the concentrations of vacancies of each type are much smaller than the concentrations of corresponding oxygen sites ($c_{Vf} \ll c_{Of}$, $c_{VR} \ll c_{OR}$ and $c_{V2R} \ll c_{O2R}$), Fermi-type correlations are negligible, and we arrive at Boltzmann statistics. In this case, the expressions for $c_{Vm}$ (11) take the simple form:

$$c_{\mathrm{Vf}} = p_{\mathrm{f}}\gamma_V c_{\mathrm{V}}, \tag{14}$$

$$c_{\mathrm{VR}} = p_{\mathrm{R}}\gamma_V c_{\mathrm{V}} e^{\frac{\Delta E_{\mathrm{VR}}}{kT}}, \tag{15}$$

$$c_{\mathrm{V2R}} = p_{\mathrm{2R}}\gamma_V c_{\mathrm{V}} e^{\frac{\Delta E_{\mathrm{V2R}}}{kT}}, \tag{16}$$

where the activity coefficient $\gamma_V$ for the Boltzmann statistics is given by

$$\gamma_{\mathrm{V}} = \left(p_{\mathrm{f}} + p_{\mathrm{R}} e^{\frac{\Delta E_{\mathrm{VR}}}{kT}} + p_{\mathrm{2R}} e^{\frac{\Delta E_{\mathrm{V2R}}}{kT}}\right)^{-1}. \tag{17}$$

### 2.3 Monte-Carlo simulation technique

The Monte Carlo modeling was conducted using a supercell containing 30×30×30 unit cells with periodic boundary conditions. Dopant concentration and temperature were varied from 0.05 to 0.5 (in steps of 0.05) and from 600 to 1500 K (in steps of 25 K), respectively.

For each combination of dopant content and temperature, the simulation started with the distribution of acceptor impurities over the B-sublattice and oxygen vacancies over the O-sublattice of the perovskite structure. Impurity ions were allocated according to the specified distribution types, and vacancies were distributed taking into account the prohibition rules for site occupancy within the adopted models. The ensemble of vacancies was then equilibrated using a sequence of random Monte Carlo steps with transition probabilities given by the Metropolis algorithm that accounts for inter-vacancy interactions within one of the adopted models (see Section 3.1).

Our calculations indicate that $10^5$ Monte Carlo steps were sufficient to achieve equilibrium for the difference in vacancy energies in neighboring sites up to ~ 0.4 eV and the temperature range of 600-1500 K. After equilibration, the simulation continued for an additional $10^4$ steps, with the concentration of vacancies, $c_{Vm}$, with different local surroundings recorded every 500 steps. To minimize the impact of random fluctuations, the resulting concentration values were obtained by averaging these measurements.

To ensure that the attained equilibrium state corresponds to the global minimum of the thermodynamic potential and to exclude possible dependence of the results on specific initial impurity configurations, the calculations were carried out repeatedly. Each simulation was performed about 10 times for every dopant content and temperature, each time starting with a new allocation of vacancies and impurities. The close agreement between repeatedly simulated $c_{Vm}(x, T)$ dependencies indicates a low probability of occurrence of the above effects during modeling.

Similar Monte Carlo simulation schemes are described in more detail in our previous works [20,40].

# 3 Results and discussion

## 3.1 Distribution of oxygen vacancies

First, we discuss the impact of trapping energies, dopant content and temperature on the distribution of vacancies over oxygen sites with different number of neighboring impurities (B-O-B, B-O-R, and R-O-R). Fig. 2 shows the concentrations of acceptor-bound vacancies, $c_{\mathrm{VR}}$ and $c_{\mathrm{V2R}}$, calculated using Eqs. (11) and (12) as a function of the trapping energy $\Delta E_{\mathrm{VR}}/kT$ and the ratio $\Delta E_{\mathrm{V2R}}/\Delta E_{\mathrm{VR}}$. Hereafter, all vacancy concentrations are normalized to their total concentration $c_{\mathrm{V}} = x/2$. As seen in Fig. 2a, $c_{\mathrm{VR}}$ attains its maximum value at high $\Delta E_{\mathrm{VR}}$ energies and low $\Delta E_{\mathrm{V2R}}/\Delta E_{\mathrm{VR}}$ ratios. An increase in $\Delta E_{\mathrm{V2R}}/\Delta E_{\mathrm{VR}}$ results in a decrease in $c_{\mathrm{VR}}$. Correspondingly, in the region of high values of $\Delta E_{\mathrm{VR}}$ and $\Delta E_{\mathrm{V2R}}/\Delta E_{\mathrm{VR}}$, $\mathrm{V_{2R}}$ vacancies become the dominant defects, as shown in Fig. 2b.

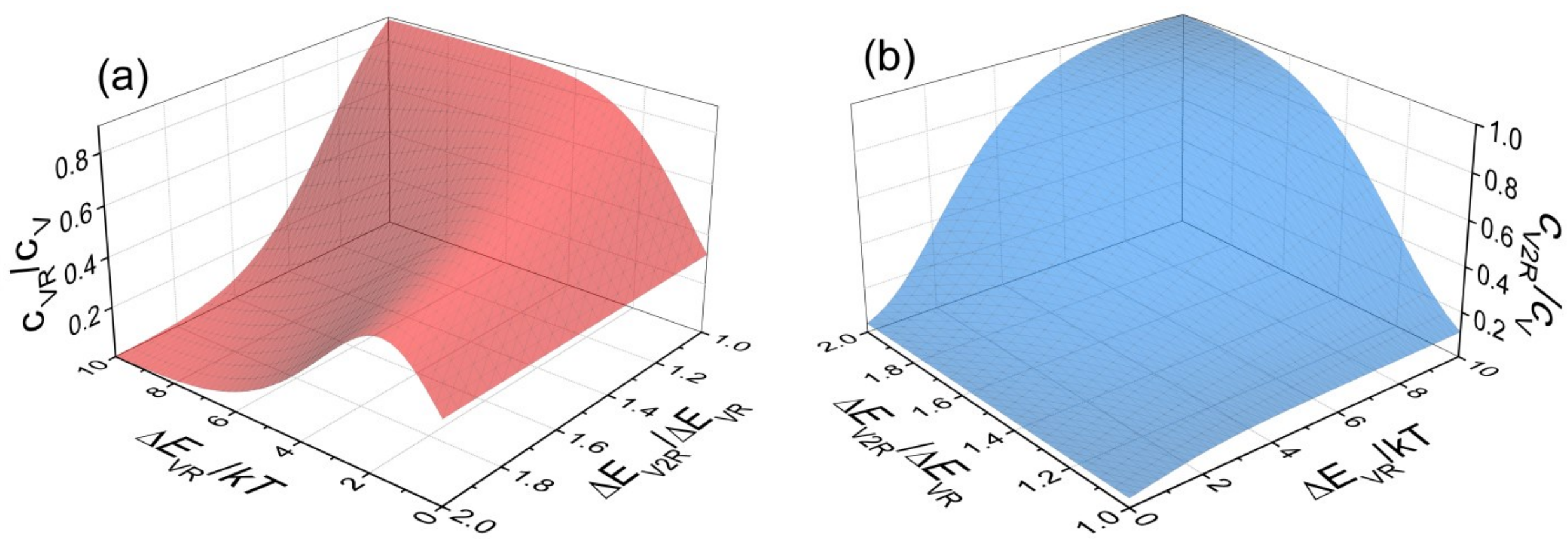

**Fig. 2.** Concentrations of oxygen vacancies in bound states near one ($c_{\mathrm{VR}}$) and two ($c_{\mathrm{V2R}}$) impurities as a function of the trapping energy $\Delta E_{\mathrm{VR}}/kT$ and the ratio $\Delta E_{\mathrm{V2R}}/\Delta E_{\mathrm{VR}}$ ($x = 0.2$). Results are presented for the model with weak (Fermi-type) inter-vacancy correlations.

Fig. 3 depicts the concentrations of vacancies in free and bound states as a function of dopant content $x$ for the typical trapping energy $\Delta E_{\mathrm{VR}} = 0.4$ eV and various $\Delta E_{\mathrm{V2R}}/\Delta E_{\mathrm{VR}}$ ratios. The Monte Carlo simulation results perfectly match the analytical calculations. With increasing $x$, the fraction of free vacancies $c_{\mathrm{Vf}}/c_{\mathrm{V}}$ decreases rapidly, so that bound vacancies become the dominant defects even at low dopant levels. The dependence of $c_{\mathrm{VR}}/c_{\mathrm{V}}$ on $x$ is non-monotonic, with its maximum shifting towards lower $x$ as $\Delta E_{\mathrm{VR}}/kT$ and/or $\Delta E_{\mathrm{V2R}}/\Delta E_{\mathrm{VR}}$ increase (for the typical trapping energies used in Fig. 3, the maximum occurs at $x < 10$ %). In contrast, $c_{\mathrm{V2R}}/c_{\mathrm{V}}$ increases monotonically with increasing $x$. At high $\Delta E_{\mathrm{V2R}}/\Delta E_{\mathrm{VR}}$ ratios and doping levels, most vacancies reside at oxygen sites surrounded by two impurities.

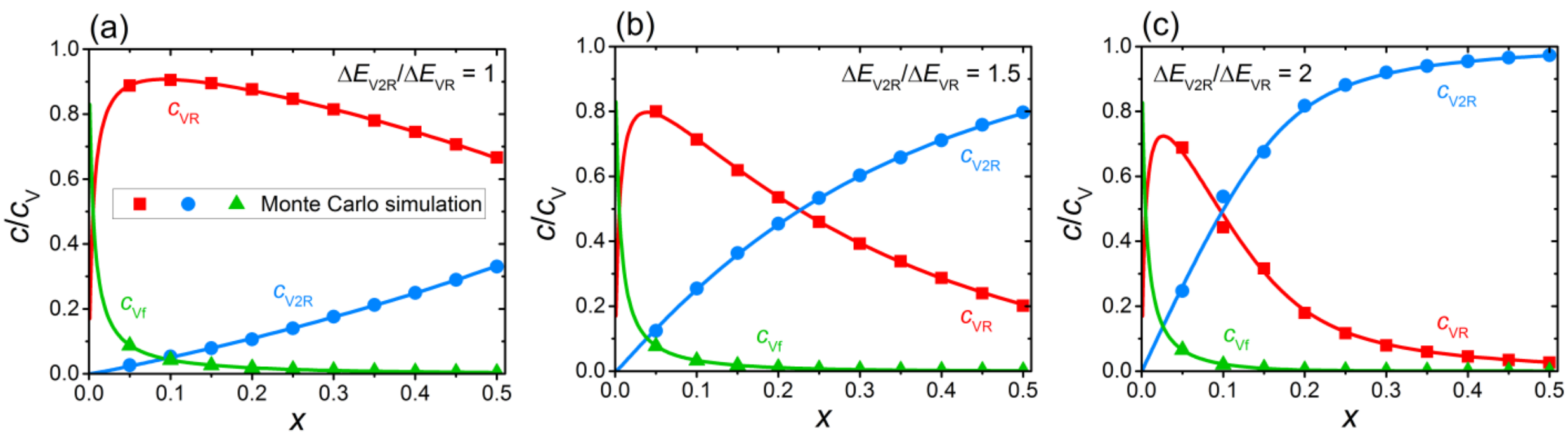

**Fig. 3.** Concentrations of acceptor-bound ($c_{\mathrm{VR}}$ and $c_{\mathrm{V2R}}$) and free ($c_{\mathrm{Vf}}$) oxygen vacancies, normalized to the total vacancy concentration $c_{\mathrm{V}} = x/2$, as a function of dopant content $x$ ($T = 1000$ K, $\Delta E_{\mathrm{VR}} = 0.4$ eV). Results are presented for three $\Delta E_{\mathrm{V2R}}/\Delta E_{\mathrm{VR}}$ ratios: (a) 1, (b) 1.5 and (c) 2. Lines and symbols correspond to the results of analytical theory and Monte Carlo simulations, respectively, for $c_{\mathrm{VR}}$ (red), $c_{\mathrm{V2R}}$ (blue) and $c_{\mathrm{Vf}}$ (green).

Fig. 4 shows the vacancy occupation probabilities $\langle n_{\mathrm{VR}}\rangle$ and $\langle n_{\mathrm{V2R}}\rangle$ (given by Eq. (9)) as a function of dopant content. The results for free vacancies are not shown because of the very low $\langle n_{\mathrm{Vf}}\rangle$ values. As expected, with decreasing temperature, the occupancy of R-O-R sites increases, and the occupancy of R-O-B sites decreases, providing $\Delta E_{\mathrm{V2R}}/\Delta E_{\mathrm{VR}} > 1$. Fig. 4 indicates that the occupation probability $\langle n_{\mathrm{VR}}\rangle$ remains small over the considered range of $x$ values ($\langle n_{\mathrm{VR}}\rangle \ll 1$), whereas the mean occupancy of sites surrounded by two impurities $\langle n_{\mathrm{V2R}}\rangle$ can be significant at low $x$ and high $\Delta E_{\mathrm{V2R}}/\Delta E_{\mathrm{VR}}$ ratios. Thus, in the latter range of parameters, Fermi-type correlations should substantially contribute to defect thermodynamics (see Sections 3.2 and 3.5).

The unusually high values of $\langle n_{\mathrm{V2R}}\rangle$ at low $x$ are explained by the low concentration of R-O-R sites and their preferential occupancy at $\Delta E_{\mathrm{V2R}}/\Delta E_{\mathrm{VR}} > 1$. The maximum on the dependence of $\langle n_{\mathrm{V2R}}\rangle$ on $x$ appears at $\Delta E_{\mathrm{V2R}}/\Delta E_{\mathrm{VR}}$ ratios exceeding a certain value $(\Delta E_{\mathrm{V2R}}/\Delta E_{\mathrm{VR}})^*$, which can be estimated using Boltzmann statistics as $\frac{kT}{\Delta E_{\mathrm{VR}}}\ln\left(2e^{\frac{\Delta E_{\mathrm{VR}}}{kT}} - 1\right)$. For $\Delta E_{\mathrm{VR}} = 0.4 \div 0.6$ eV and $T = 600 \div 1200$ K, we obtain $(\Delta E_{\mathrm{V2R}}/\Delta E_{\mathrm{VR}})^* = 1.1 \div 1.2$. With increasing the ratio $\Delta E_{\mathrm{V2R}}/\Delta E_{\mathrm{VR}}$ above $(\Delta E_{\mathrm{V2R}}/\Delta E_{\mathrm{VR}})^*$, the occupancy $\langle n_{\mathrm{V2R}}\rangle$ increases and approaches unity at $\Delta E_{\mathrm{V2R}}/kT \gtrsim 2\Delta E_{\mathrm{VR}}/kT \gg 1$, when nearly all vacancies are surrounded by two impurities. We can estimate the region of $x$ values, in which $\langle n_{\mathrm{V2R}}\rangle$ can be high, using the condition $c_{\mathrm{V}} = x/2 \gtrsim 3p_{2\mathrm{R}}$, yielding $x \lesssim 0.17$. The concentration $x_0$, at which $\langle n_{\mathrm{V2R}}\rangle$ attains its maximum, shifts to higher $x$ with decreasing $\Delta E_{\mathrm{VR}}/kT$ and/or $\Delta E_{\mathrm{V2R}}/\Delta E_{\mathrm{VR}}$. At realistic energy parameters $\Delta E_{\mathrm{VR}} = 0.4$ eV and $\Delta E_{\mathrm{V2R}}/\Delta E_{\mathrm{VR}} = 1.5$, the $x_0$ value falls within the range $0 < x_0 < 0.07$ at temperatures between 600 and 1200 K.

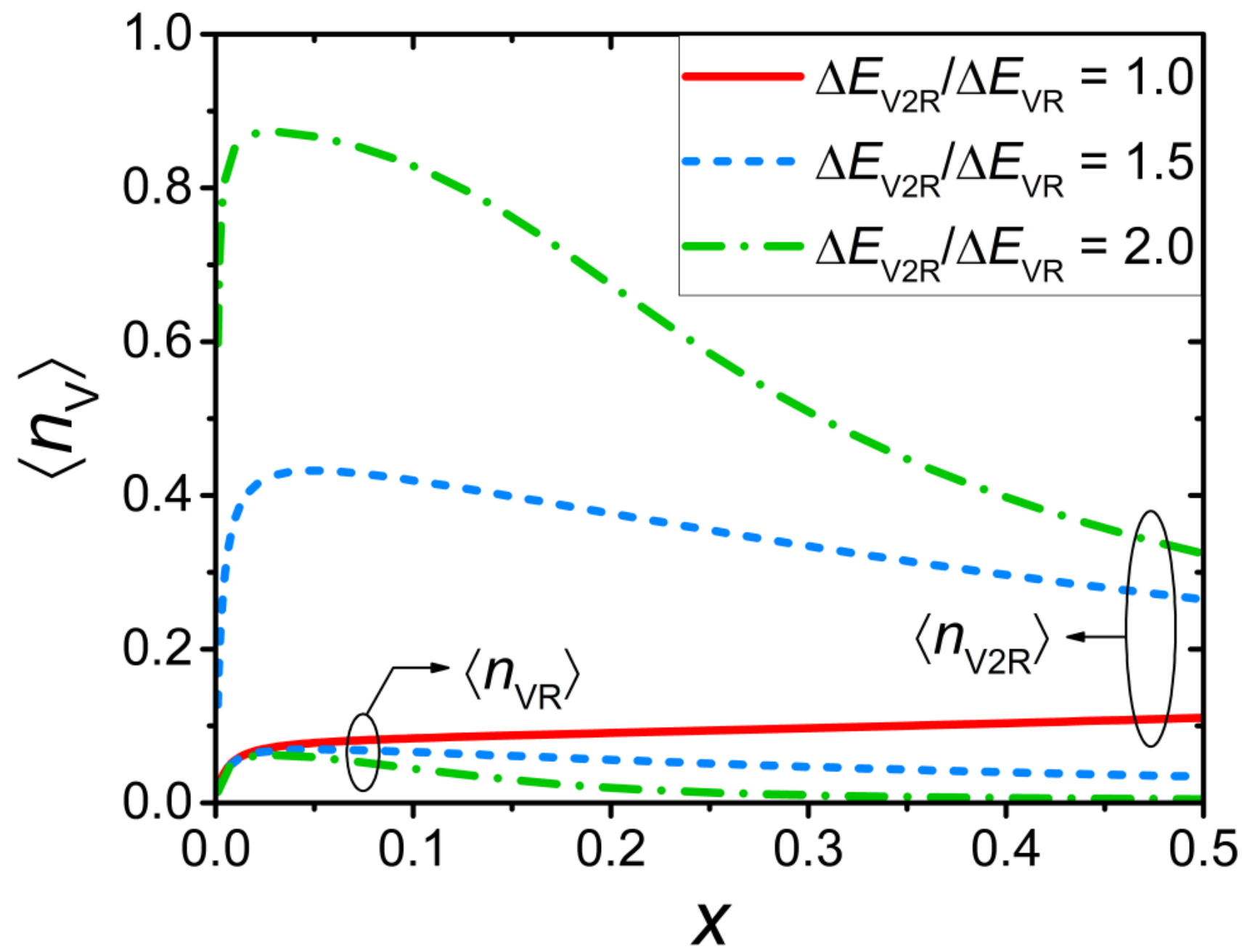

**Fig. 4.** Occupation probabilities (Eq. (9)) for oxygen vacancies bound near one ($\langle n_{\mathrm{VR}} \rangle$) and two ($\langle n_{\mathrm{V2R}} \rangle$) impurities as a function of dopant content $x$ at different $\Delta E_{\mathrm{V2R}}/\Delta E_{\mathrm{VR}}$ ratios ($T = 1000$ K, $\Delta E_{\mathrm{VR}} = 0.4$ eV).

### 3.2 Correlations

Correlations between oxygen vacancies can play a significant role in their distribution, particularly in heavily doped perovskites. We explore the effects of these correlations using analytical calculations and Monte-Carlo simulations for the proposed model (see Section 2.1). The analytical approach accounts only for weak (Fermi-type) correlations, while the Monte Carlo modeling incorporates both weak and strong (inter-site) correlations.

Fig. 5 illustrates the impact of correlations on vacancy distribution over oxygen sites with different impurity surroundings. The correlations lead to an increase in the concentration of $\mathrm{V_R}$ vacancies, $c_{\mathrm{VR}}$, and a decrease in the concentration of $\mathrm{V_{2R}}$ vacancies, $c_{\mathrm{V2R}}$. This effect becomes more pronounced at higher values of the ratio $\Delta E_{\mathrm{V2R}}/\Delta E_{\mathrm{VR}}$. The reason for such an opposite trend lies in the significantly different occupation probabilities of $\mathrm{V_R}$ and $\mathrm{V_{2R}}$ vacancies (see Fig. 4). Since $\langle n_{\mathrm{V2R}} \rangle$ is relatively high, inter-vacancy correlations noticeably reduce the number of available oxygen sites, thereby decreasing $c_{\mathrm{V2R}}$.

The effect of inter-vacancy interactions is most pronounced at low temperatures, as shown in Fig. 5b. When inter-site vacancy repulsion is neglected, almost all vacancies occupy the lowest energy states (R-O-R sites). Accounting for this repulsion leads to a decrease in $c_{\mathrm{V2R}}$ and an increase in $c_{\mathrm{VR}}$. In this case, the fraction of $\mathrm{V_R}$ vacancies is significant even at low temperatures. The effect of strong correlations diminishes with decreasing dopant concentration. The change in temperature mainly results in the redistribution of vacancies between B-O-R and R-O-R sites, as the fraction of free vacancies $c_{\mathrm{Vf}}/c_{\mathrm{V}}$ is small under the considered conditions.

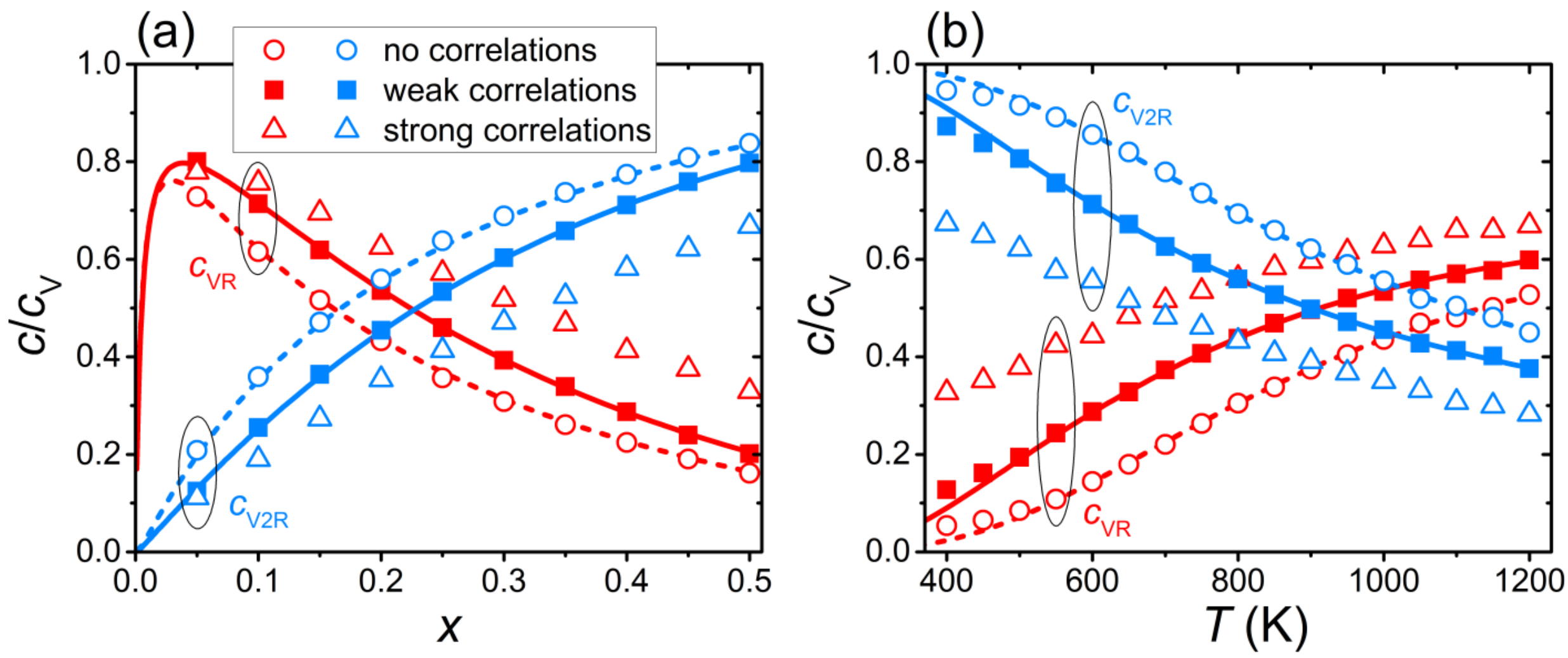

**Fig. 5.** Concentrations of oxygen vacancies bound near one ($c_{VR}$) and two ($c_{V2R}$) acceptor impurities as a function of (a) dopant content $x$ ($T = 1000$ K) and (b) temperature $T$ ($x = 0.2$). Results are presented for $\Delta E_{VR} = 0.4$ eV and $\Delta E_{V2R}/\Delta E_{VR} = 1.5$. Symbols indicate Monte Carlo results for three models: no correlations (open circles), weak correlations (closed squares) and strong correlations (open triangles). Dashed and solid lines represent analytically calculated dependencies for the models without correlations and with weak inter-vacancy correlations, respectively.

To analyze the effect of inter-vacancy correlations on local structure, we consider the short-range order parameter $\alpha$, which characterizes deviations in the local distribution of host ions and/or defects from an uncorrelated arrangement. We define $\alpha$ following Ref. [41] and express it as

$$\alpha = 1 - Z/Z_0, \tag{18}$$

where $Z$ and $Z_0$ are the first-nearest-neighbor (1NN) coordination numbers with and without inter-vacancy correlations.

Let us consider the average coordination numbers of oxygen vacancies around acceptor impurities $Z_{\text{R-V}}$ and host B cations $Z_{\text{B-V}}$. For the uniform dopant distribution, these coordination numbers are given by

$$Z_{\mathrm{R-V}} = 6[x\langle n_{\mathrm{V2R}}\rangle + (1-x)\langle n_{\mathrm{VR}}\rangle], \tag{19}$$

$$Z_{\mathrm{B-V}} = 6[x\langle n_{\mathrm{VR}}\rangle + (1-x)\langle n_{\mathrm{Vf}}\rangle]. \tag{20}$$

Using the expressions for $p_{\text{m}}$ (Eq. (7)) and the relationship between occupation probabilities $\langle n_{\text{Vm}}\rangle$ and concentrations $c_{\text{Vm}}$ (Eq. (11)), from Eqs. (19) and (20) we obtain

$$Z_{\mathrm{R-V}} = \frac{c_{\mathrm{VR}} + 2c_{\mathrm{V2R}}}{x}, \tag{21}$$

$$Z_{\mathrm{B-V}} = \frac{x - c_{\mathrm{VR}} - 2c_{\mathrm{V2R}}}{1-x}. \tag{22}$$

The calculated short-range order parameters $\alpha_{\text{R-V}}$ and $\alpha_{\text{B-V}}$ as a function of dopant content are given in Fig. 6. The contribution of weak correlations exhibits a non-monotonic dependence on $x$, the shape of which is governed by the behavior of the occupation probabilities varying non-monotonically with $x$ (see Fig. 4).

Specifically, high values of $\langle n_{V2R} \rangle$ in a narrow doping range lead to large deviations of vacancy concentrations from their values in the absence of correlations, which in turn results in large deviations of the short-range order parameters from zero. At low $x$, $\alpha_{R\text{-}V}$ and $\alpha_{B\text{-}V}$ can be largely determined by weak correlations, while at high $x$, strong correlations play a dominant role.

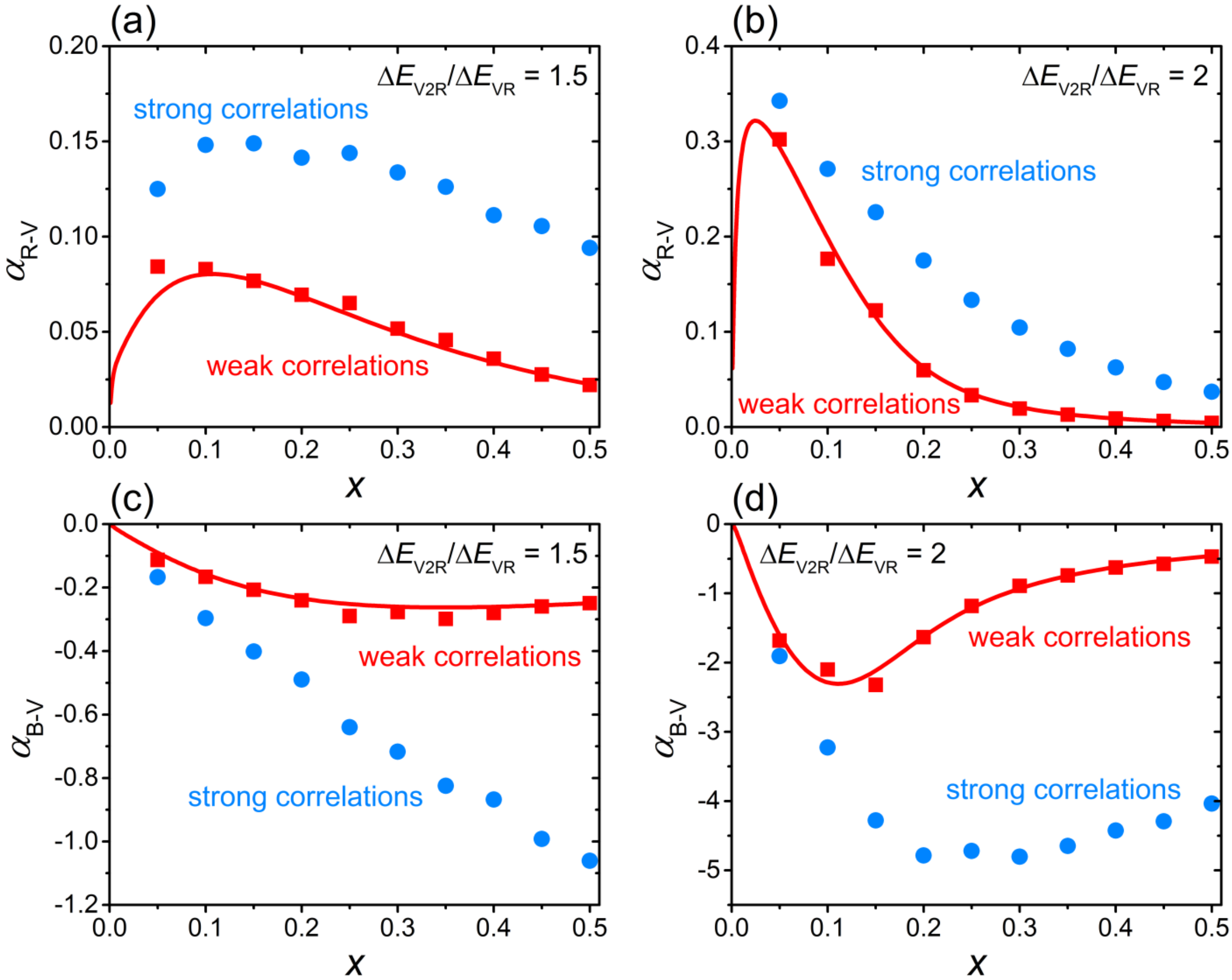

**Fig. 6.** Short-range order parameters for the distribution of vacancies around (a,b) impurities ($\alpha_{R\text{-}V}$) and (c,d) host B cations ($\alpha_{B\text{-}V}$) as a function of dopant content $x$ ($T = 1000$ K). Results are presented for $\Delta E_{VR} = 0.4$ eV and two $\Delta E_{V2R}/\Delta E_{VR}$ ratios: (a,c) 1.5 and (b,d) 2. Lines and symbols indicate the results of analytical theory and Monte Carlo simulations for weak (red lines and squares) and strong (blue circles) correlations.

### 3.3 Coordination numbers

Studying coordination numbers is useful for understanding the effect of inter-defect interactions on the local structure of oxides. Fig. 7 depicts the average coordination numbers $Z_{B\text{-}O}$ and $Z_{R\text{-}O}$ of host B cations and impurities R, respectively, surrounded by 1NN oxygen ions as a function of $x$. These coordination numbers are calculated as $Z_{B-O} = 6 - Z_{B-V}$ and $Z_{R-O} = 6 - Z_{R-V}$ using Eqs. (19) and (20). At uniform defect distribution, when inter-defect interactions are absent, $Z_{B\text{-}O}$ and $Z_{R\text{-}O}$ reduce to $6 - x$ (black dashed line in Fig. 7a).

With increasing $x$, and accordingly the concentration of vacancies in the oxide, both $Z_{\text{B-O}}$ and $Z_{\text{R-O}}$ decrease. However, at high values of $\Delta E_{\text{V2R}}/\Delta E_{\text{VR}}$, most vacancies are surrounded by two impurities, and $Z_{\text{B-O}}$ remains close to 6 (Fig. 7a). In the latter case, an impurity on average traps one vacancy and $Z_{\text{R-O}}$ approaches 5.

To assess the effect of vacancy-impurity interactions, we consider the short-range order parameter $\alpha^*$, defined analogously to Eq. (18), but with $Z_0$ being replaced by $Z_0^*$ – the coordinating number in the absence of inter-defect interactions. This parameter characterizes deviations from the uniform defect distribution. As seen in Fig. 7, the average coordination numbers $Z_{\text{B-O}}$ and $Z_{\text{R-O}}$ lie, respectively, above and below their values at uniform distribution, leading to $\alpha^*_{\text{B-O}} < 0$ and $\alpha^*_{\text{R-O}} > 0$. Note that $\alpha^*_{\text{B-O}}$ decreases monotonically with increasing $x$, while $\alpha^*_{\text{R-O}}(x)$ exhibits a maximum at moderate $x$. As $\Delta E_{\text{V2R}}/\Delta E_{\text{VR}}$ increases, the defect distribution becomes increasingly non-uniform (Fig. 7b).

The Monte Carlo simulation results in Fig. 7 are shown for the cases of weak and strong correlations. Although the coordination numbers remain close in both models, accounting for inter-site vacancy interactions slightly increases $Z_{\text{R-O}}$ due to the reduced number of vacancies surrounding each impurity.

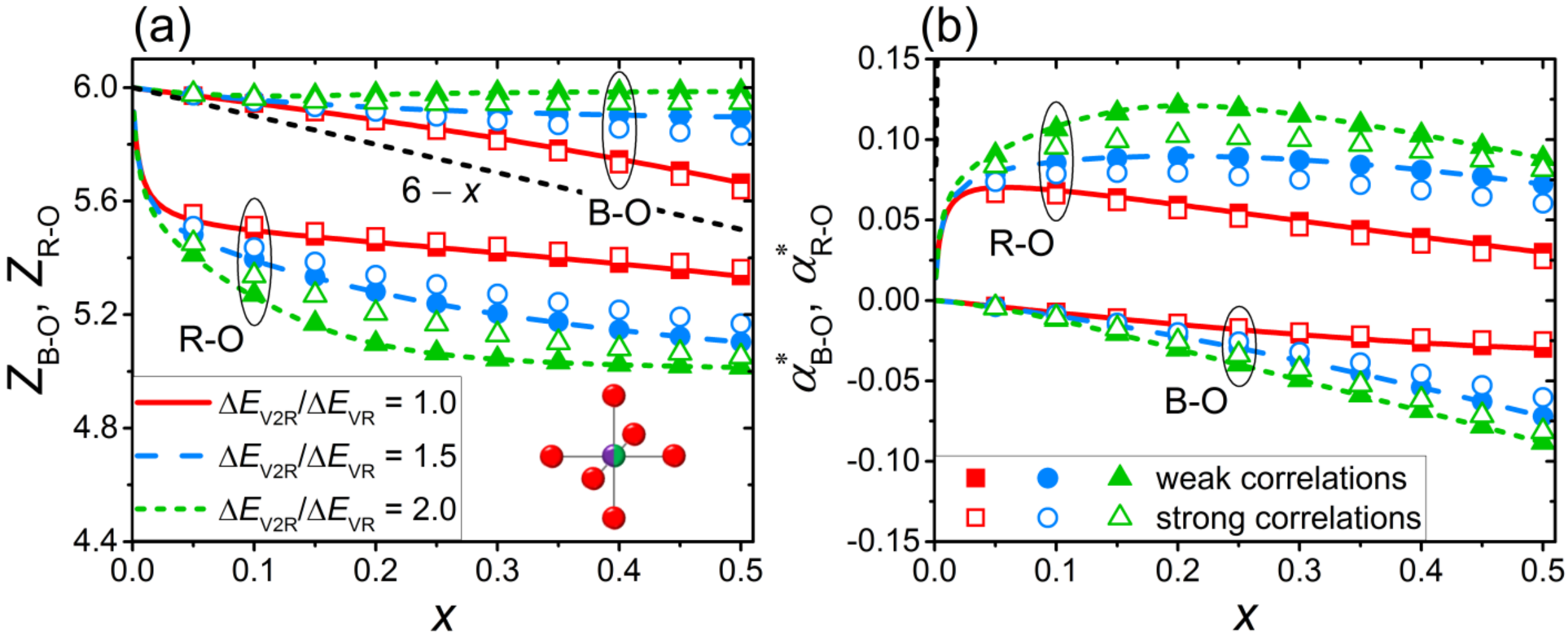

**Fig. 7.** (a) Coordination numbers of B cations ($Z_{\text{B-O}}$) and impurities R ($Z_{\text{R-O}}$) surrounded by oxygen ions and (b) the corresponding short-range order parameters $\alpha^*_{\text{B-O}}$ and $\alpha^*_{\text{R-O}}$ as a function of dopant content $x$ ($T$ = 1000 K, $\Delta E_{\text{VR}}$ = 0.4 eV). Results are provided for different $\Delta E_{\text{V2R}}/\Delta E_{\text{VR}}$ ratios. Colored lines represent analytically calculated dependencies for the model with weak correlations. Closed and open symbols correspond to Monte Carlo results for the model with weak and strong correlations, respectively. Dashed black lines correspond to the uniform distribution of defects. The fragment of the perovskite structure illustrating the nearest-neighbor coordination of B/R cations is given in (a), where oxygen and B/R ions are indicated by red and violet-green spheres, respectively.

We now proceed to analyze the behavior of the average coordination number of acceptor impurities R around oxygen vacancies V, given by

$$Z_{\mathrm{V-R}} = 2Z_{\mathrm{R-V}} = \frac{c_{\mathrm{VR}}+2c_{\mathrm{V2R}}}{c_{\mathrm{V}}}. \qquad (23)$$

Fig. 8a shows that $Z_{\mathrm{V\text{-}R}}$ increases with increasing $x$ and is higher than its value at uniform distribution of defects ($2x$). Correspondingly, the parameter $\alpha^*_{\mathrm{V-R}}$ is negative (Fig. 8b), indicating the attractive interaction between vacancies and impurities. The effect of strong correlations results in the slightly reduced $Z_{\mathrm{V\text{-}R}}$ values (Fig. 8a). As shown in Fig. 8b, the largest deviation from the uniform distribution occurs at low $x$. With increasing $x$, $\alpha^*_{\mathrm{V-R}}$ increases and gradually approaches zero at high doping levels. Increasing the ratio $\Delta E_{\mathrm{V2R}}/\Delta E_{\mathrm{VR}}$ leads to an increase in the concentration of R-V-R complexes and, as a result, $Z_{\mathrm{V\text{-}R}}$ approaches 2.

A comparison between the short-range order parameters describing the trapping ($\alpha^*_{\mathrm{R-V}}$, $\alpha^*_{\mathrm{B-V}}$, $\alpha^*_{\mathrm{B-O}}$) and correlation ($\alpha_{\mathrm{R\text{-}V}}$, $\alpha_{\mathrm{B\text{-}V}}$, $\alpha_{\mathrm{B\text{-}O}}$) effects shows that the former significantly exceeds the latter for most local coordination environments, indicating the predominant role of vacancy-impurity interactions.

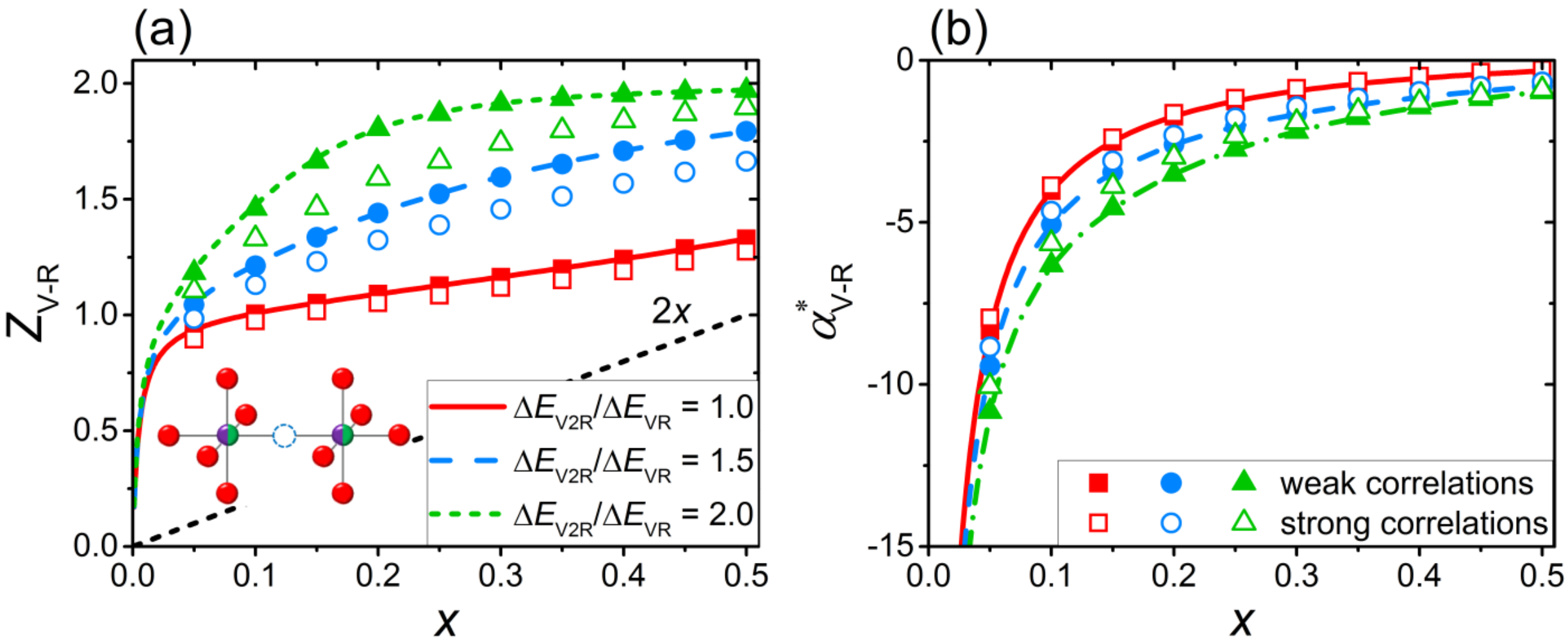

**Fig. 8.** (a) Coordination number $Z_{\mathrm{V\text{-}R}}$ of oxygen vacancies V surrounded by acceptor impurities R and (b) the corresponding short-range order parameter $\alpha^*_{\mathrm{V-R}}$ as a function of dopant content $x$ ($T = 1000$ K, $\Delta E_{\mathrm{VR}} = 0.4$ eV). Results are provided for different $\Delta E_{\mathrm{V2R}}/\Delta E_{\mathrm{VR}}$ ratios. Colored lines indicate analytical theory results for the model with weak correlations. Closed and open symbols represent Monte Carlo simulation results for the model with weak and strong correlations, respectively. The dashed black line corresponds to the uniform distribution of vacancies over oxygen sites. The fragment of the perovskite structure illustrating the nearest-neighbor coordination of an oxygen vacancy is given in (a). Oxygen ions, vacancies and B/R cations are indicated by red, white and violet-green spheres, respectively.

### 3.4 Non-uniform distribution of impurities

Our considerations thus far have focused on the problem at uniform dopant distribution in oxides. The actual distribution, however, can deviate from uniform due to a sample preparation method. In this section, we discuss the implications of a non-uniform distribution of acceptor impurities for the defect and local structure of the studied $AB_{1-x}R_xO_{3-\delta}$ perovskites. As an example, we use a frozen high-temperature dopant arrangement corresponding to the equilibrium structure obtained via combined DFT and Monte Carlo simulations [30]. Fig. 9a presents the data extracted from Ref. [30] on the average Y-Y coordination at 1NN sites in Y-doped $BaZrO_3$ at a typical sintering temperature of 1900 K. It is seen that the probability of finding a Y ion at a B-site nearest to another Y ion is noticeably lower than expected for the uniform dopant distribution.

To incorporate this non-uniform dopant distribution into our theory, we extract the necessary probabilities $p_R$, $p_{2R}$ and $p_f$ from the Y-Y coordination data [30] (see Fig. 9a). This is achieved by determining from the data [30] the conditional probabilities $p(j|i)$, which describe the likelihood of finding a cation of type $j$ at a B-site adjacent to a cation of type $i$ ($i, j$ = B, R). These probabilities are related to $p_m$ as follows:

$$p_R = x\big(1 - p(R|R)\big) + (1 - x)p(R|B), \tag{24a}$$

$$p_{2R} = xp(R|R), \tag{24b}$$

$$p_f = (1 - x)\big(1 - p(R|B)\big). \tag{24c}$$

For the uniform dopant distribution, we have $p(R|R) = p(R|B) = x$.

The R-R coordination number at a 1NN site is given by $6p(R|R)$. Thus, using the data from Ref. [30], we determine $p(R|R)$ and $p(R|B)$, allowing us to calculate the desired probabilities $p_m$ for each dopant content. Fig. 9b demonstrates that $p_R$ and $p_f$, calculated using Eqs. (24a) and (24c), are close to their values ($p_R^u$ and $p_f^u$) at uniform doping, whereas $p_{2R}$ (24b) is significantly lower than $p_{2R}^u$.

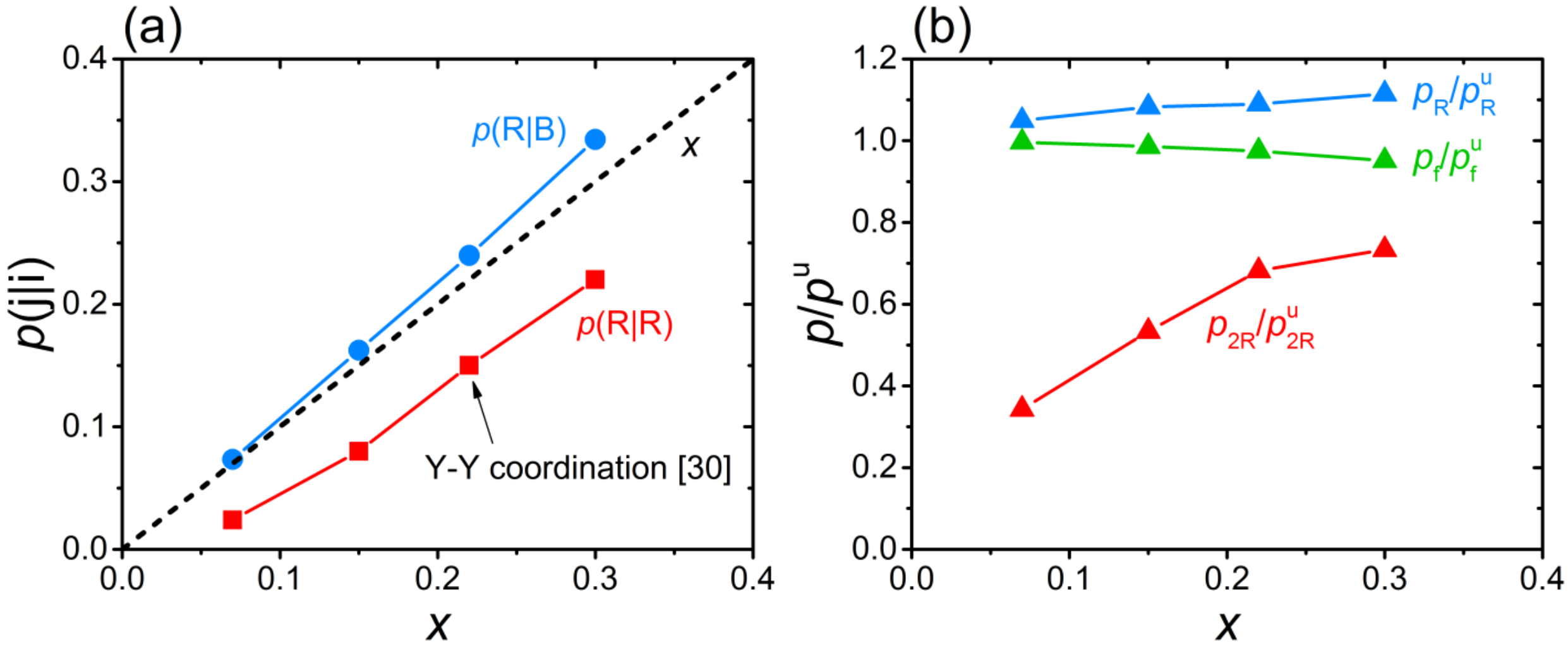

**Fig. 9.** (a) Conditional probabilities $p(R|R)$ and $p(R|B)$ extracted from the Y-Y coordination data at 1NN sites in Y-doped $BaZrO_3$ at ~ 1900 K obtained using DFT and Monte Carlo simulation [30]. The dashed black line corresponds to the probability for the uniform impurity distribution. (b) The ratio of the probabilities for a non-uniform dopant distribution ($p_m$) to those at uniform doping ($p_m^u$), plotted as a function of dopant content $x$. Symbols indicate these ratios calculated from the data given in (a) using Eqs. (24).

With the obtained probabilities $p_{\mathrm{R}}$, $p_{2\mathrm{R}}$ and $p_{\mathrm{f}}$, we can compute the defect concentrations using Eq. (11). Fig. 10a illustrates that the employed dopant distribution results in a decrease in the concentration of R-V-R complexes ($c_{\mathrm{V2R}}$) and an increase in the concentration of R-V-B complexes ($c_{\mathrm{VR}}$) compared to the uniform distribution.

The coordination numbers $Z_{\mathrm{R\text{-}V}}$ (19) and $Z_{\mathrm{B\text{-}V}}$ (20) for the considered non-uniform dopant distribution can be found using the determined probabilities $p(j|i)$ and mean occupancy numbers $\langle n_{\mathrm{m}} \rangle$:

$$Z_{\mathrm{R-V}} = 6[p(\mathrm{R}|\mathrm{R})\langle n_{\mathrm{V2R}} \rangle + p(\mathrm{B}|\mathrm{R})\langle n_{\mathrm{VR}} \rangle], \tag{25}$$

$$Z_{\mathrm{B-V}} = 6[p(\mathrm{R}|\mathrm{B})\langle n_{\mathrm{VR}} \rangle + p(\mathrm{B}|\mathrm{B})\langle n_{\mathrm{Vf}} \rangle]. \tag{26}$$

Using Eqs. (24), these expressions reduce to the previously derived formulas (21) and (22).

Fig. 10b shows the ratio of the coordination numbers at non-uniform impurity distribution to their values at uniform doping. The coordination numbers $Z_{\mathrm{R\text{-}O}}$ and $Z_{\mathrm{B\text{-}O}}$ computed using Eqs. (25) and (26) are virtually the same as those for the uniform dopant distribution (the results in Fig. 10b are shown only for $Z_{\mathrm{R\text{-}O}}$). At the same time, the employed non-uniform dopant distribution slightly reduces the average number of Y ions surrounding V, especially at high $\Delta E_{\mathrm{V2R}}/\Delta E_{\mathrm{VR}}$ ratios. This reduction arises from a decreased probability of the formation of R-R complexes (Fig. 10b).

To obtain a more general assessment of the influence of non-uniform dopant distribution, we calculated the coordination numbers at realistic trapping energies in a wide range of temperatures and dopant concentrations, varying the probability $p(\mathrm{R}|\mathrm{R})$ by $\pm$ 50 % relative to its value at uniform distribution. The results indicate that the maximum deviation from the values at uniform doping does not exceed ~ 2 %, ~ 5 % and ~ 25% for $Z_{\mathrm{B\text{-}O}}$, $Z_{\mathrm{R\text{-}O}}$ and $Z_{\mathrm{V\text{-}R}}$, respectively.

Note that the obtained results should be considered as an illustration of the possible impact of non-uniform dopant distribution rather than as general predictions for the studied perovskites. In real oxides, dopant distribution depends on the synthesis method and sample preparation history.

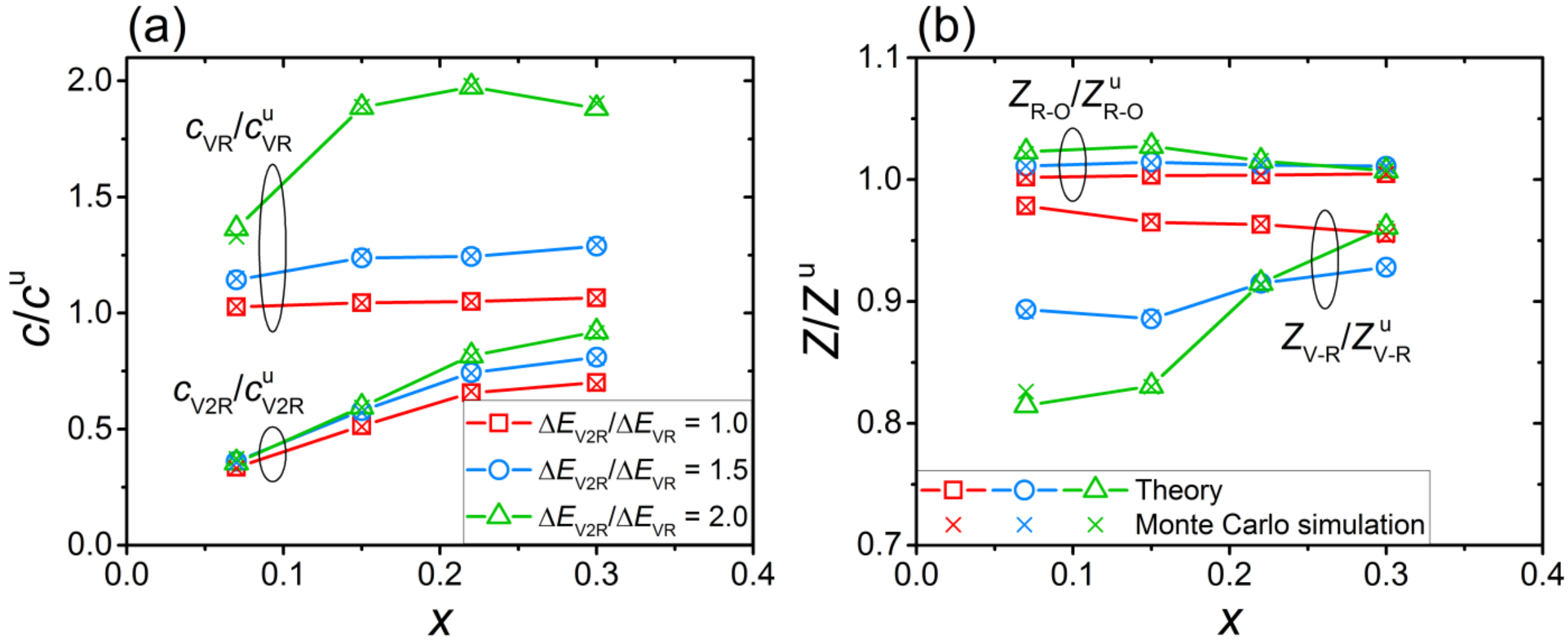

**Fig. 10.** Ratios of the (a) oxygen vacancy concentrations ($c_{VR}$ and $c_{V2R}$) and (b) coordination numbers ($Z_{R\text{-}O}$ and $Z_{V\text{-}R}$) at non-uniform impurity distribution to their values at uniform doping ($c^u$ and $Z^u$) as a function of dopant content $x$ ($T = 1000$ K, $\Delta E_{VR} = 0.4$ eV). Open symbols and crosses indicate the results of our analytical theory and Monte Carlo modeling, respectively. The dependencies are presented for the non-uniform dopant distribution corresponding to a frozen high-temperature (~ 1900 K) cation allocation obtained via DFT and Monte Carlo simulations [30].

### 3.5 Oxidation

In this section, we examine the influence of inter-defect interactions on the oxidation of wide-band-gap $AB_{1-x}R_xO_{3-\delta}$ perovskites without mixed-valence cations. In such oxides, the states of electron holes are derived mostly from O2p orbitals [25-27,42]. Typically, small hole polarons in wide-band-gap oxides are energetically more favorable than free valence-band holes [25,42-45]. The holes are mainly localized on oxygen ions adjacent to impurities, forming acceptor-bound small hole polarons [23,25,42,45,46].

As we showed earlier [23,46], if the acceptor-bound hole states are located deep enough in the band gap, the oxidation reaction can be exothermic, and the concentration of bound holes can be considerable, and even comparable to that of ionic defects. In most cases, however, one can expect endothermic oxidation of the studied oxides and, accordingly, the hole concentration $c_h$ much lower than $c_V$. Assuming the condition $c_h << c_V$, we can neglect the effect of oxidation on the oxygen content and hence $c_V = x/2$.

Despite the fact that $c_h << c_V$, the hole conductivity $\sigma_h$ can be comparable to or even exceed the oxygen-ion conductivity in perovskites under consideration due to high hole mobility (see, e.g., [18,23,46,47]). As we previously demonstrated, the hole transport data in these oxides can be described by both band-like transfer and small polaron hopping models, owing to the small activation energy of hopping [23,46]. In this study, for definiteness, we consider holes as small polarons (consideration of band-like holes instead of small polarons does not affect the results on oxidation).

The oxidation reaction and the corresponding oxide-gas equilibrium condition can be written as

$$0.5\mathrm{O}_{2(\mathrm{g})} \Leftrightarrow \mathrm{O}^{2-} + 2\mathrm{h}^+, \qquad (27)$$

$$0.5\mu_{\mathrm{O2}} = 2\mu_{\mathrm{h}} - \mu_{\mathrm{V}}, \qquad (28)$$

where $\mu_{O2}$ is the chemical potential of oxygen molecules in the gas phase, defined as $\mu_{O2} = \mu_{O2}^0 + kT \ln p_{O2}$ ($\mu_{O2}^0$ is the standard chemical potential and $p_{O2}$ is the oxygen partial pressure); $\mu_h$ and $\mu_V$ are the chemical potentials of holes and vacancies, respectively.

Given the low concentration of holes, their chemical potential can be written as

$$\mu_{\mathrm{h}} = \mu_{\mathrm{h}}^0 + kT \ln c_{\mathrm{h}}, \qquad (29)$$

where $c_h$ is the concentration of hole carriers and $\mu_h^0$ is their standard chemical potential.

Using Eqs. (13), (28) and (29), we can write the mass action law for the oxidation reaction (27) as follows:

$$K_{\mathrm{ox}} \equiv K_{\mathrm{ox}}^{0} K_{\mathrm{ox}}^{\mathrm{int}} = \frac{3c_{\mathrm{h}}^{2}}{c_{\mathrm{V}}\sqrt{p_{\mathrm{O2}}}}, \tag{30}$$

where we introduce the effective oxidation constant $K_{\mathrm{ox}}$, which includes the equilibrium constant $K_{\mathrm{ox}}^{0}$ for an ideal defect solution and the non-ideality factor $K_{\mathrm{ox}}^{\mathrm{int}}$. The components $K_{\mathrm{ox}}^{0}$ and $K_{\mathrm{ox}}^{\mathrm{int}}$ are determined as

$$K_{\mathrm{ox}}^{0} = \exp\left(\frac{E_{\mathrm{Vf}} + 2\mu_{\mathrm{h}}^{0} + 0.5\mu_{\mathrm{O2}}^{0}}{kT}\right) \tag{31}$$

and

$$K_{\mathrm{ox}}^{\mathrm{int}} = \gamma_{\mathrm{V}}, \tag{32}$$

where $\gamma_{\mathrm{V}}$ is the activity coefficient of vacancies, defined by Eq. (13) in Section 2.2.

In the case of Boltzmann statistics, $K_{\mathrm{ox}}^{\mathrm{int}} = \gamma_{\mathrm{V}}$ can be explicitly expressed by Eq. (17) in terms of the trapping energies ($\Delta E_{\mathrm{VR}}$ and $\Delta E_{\mathrm{V2R}}$) and probabilities $p_{\mathrm{m}}$. The attractive interaction between vacancies and impurities ($\Delta E_{\mathrm{VR}}, \Delta E_{\mathrm{V2R}} > 0$) yields $\gamma_{\mathrm{V}} < 1$ (see Eq. (17)), which means that the trapping effect reduces the oxidation constant $K_{\mathrm{ox}}$, and, correspondingly, the hole concentration.

The non-ideality factor $K_{\mathrm{ox}}^{\mathrm{int}}$ as a function of dopant content is plotted in Fig. 11a. To assess the impact of Fermi-type correlations, we also plotted the ratio of $\gamma_{\mathrm{V}}$ to its value calculated using Boltzmann statistics (Eq. (17)). As shown in Fig. 11, these correlations have a small effect on oxidation, with the exception of a narrow range of parameters characterized by low temperatures, large $\Delta E_{\mathrm{V2R}}/\Delta E_{\mathrm{VR}}$ ratios and moderate dopant concentrations ($x < 0.2$). This range of parameters corresponds to the high occupancy $\langle n_{\mathrm{V2R}} \rangle$ of R-O-R sites (see Section 3.1). In contrast, vacancy trapping strongly influences oxidation, with the magnitude of this effect varying significantly with energy parameters, dopant content and temperature. In particular, the contribution of trapping to $K_{\mathrm{ox}}$ increases with increasing $x$ and decreasing $T$, as illustrated in Fig. 11a. Larger $\Delta E_{\mathrm{V2R}}/\Delta E_{\mathrm{VR}}$ ratios enhance both trapping and correlation effects.

To find the contribution of inter-defect interactions to the oxidation enthalpy, we use the standard relation $\Delta H_{\mathrm{ox}}^{\mathrm{int}} = kT^{2} d \ln K_{\mathrm{ox}}^{\mathrm{int}} / dT$. The dependence of $\Delta H_{\mathrm{ox}}^{\mathrm{int}}$ on $x$ is shown in Fig. 11b. The positive sign of $\Delta H_{\mathrm{ox}}^{\mathrm{int}}$ is explained by the fact that the trapping effect reduces the formation energy of oxygen vacancies ($\Delta E_{\mathrm{VR}}$, $\Delta E_{\mathrm{V2R}} > 0$), rendering oxidation less favorable. When the majority of vacancies occupy R-O-B sites, $\gamma_{\mathrm{V}} \sim \exp(-\Delta E_{\mathrm{VR}}/kT)$ and therefore $\Delta H_{\mathrm{ox}}^{\mathrm{int}} \approx \Delta E_{\mathrm{VR}}$. Similarly, when the majority of vacancies are surrounded by two impurity ions, $\gamma_{\mathrm{V}} \sim \exp(-\Delta E_{\mathrm{V2R}}/kT)$ and $\Delta H_{\mathrm{ox}}^{\mathrm{int}}$ approaches $\Delta E_{\mathrm{V2R}}$ at high $x$. These scenarios, illustrated in Fig. 11b, occur in the regions of low ($x \lesssim 0.1$) and high ($x > 0.2$) dopant concentrations, respectively, at reasonably large $\Delta E_{\mathrm{V2R}}/\Delta E_{\mathrm{VR}}$ ratios and low temperatures. At moderate ratios ($\Delta E_{\mathrm{V2R}}/\Delta E_{\mathrm{VR}} \sim 1.5$) and/or higher temperatures, the first plateau ($\Delta H_{\mathrm{ox}}^{\mathrm{int}} \approx \Delta E_{\mathrm{VR}}$) on the dependence $\Delta H_{\mathrm{ox}}^{\mathrm{int}}(x)$ disappears.

For the region of parameters, where the effect of Fermi-type correlations is negligible (see above), we can obtain a good estimate of $\Delta H_{\mathrm{ox}}^{\mathrm{int}}$ using Eq. (17):

$$\Delta H_{\mathrm{ox}}^{\mathrm{int}} = \frac{\Delta E_{\mathrm{VR}} p_{\mathrm{R}} e^{\frac{\Delta E_{\mathrm{VR}}}{kT}} + \Delta E_{\mathrm{V2R}} p_{\mathrm{2R}} e^{\frac{\Delta E_{\mathrm{V2R}}}{kT}}}{p_{\mathrm{f}} + p_{\mathrm{R}} e^{\frac{\Delta E_{\mathrm{VR}}}{kT}} + p_{\mathrm{2R}} e^{\frac{\Delta E_{\mathrm{V2R}}}{kT}}}. \qquad (33)$$

If $\Delta E_{\mathrm{VR}} = \Delta E_{\mathrm{V2R}} = \Delta E_{\mathrm{V}}$, Eq. (33) simplifies to

$$\Delta H_{\mathrm{ox}}^{\mathrm{int}} = \frac{\Delta E_{\mathrm{V}}}{1 + \frac{p_{\mathrm{f}}}{p_{\mathrm{R}} + p_{\mathrm{2R}}} e^{-\frac{\Delta E_{\mathrm{V}}}{kT}}}, \qquad (34)$$

which was previously obtained in Ref. [23] for low dopant concentrations. In this case, the oxidation enthalpy has only one limit, $\Delta H_{\mathrm{ox}}^{\mathrm{int}} \approx \Delta E_{\mathrm{V}}$, attained at high $x$ and $\Delta E_{\mathrm{V}}/kT$.

Fig. 11 also shows the equilibrium constant and oxidation enthalpy calculated for the frozen non-uniform dopant distribution, corresponding to the high-temperature defect configuration obtained in Ref. [30] (see Section 3.4). Although the results for the uniform and non-uniform distributions slightly differ, the overall trends in the behavior of the thermodynamic parameters of the oxidation reaction are similar.

The considered inter-defect interactions significantly affect the position of the Fermi level (defined as $\varepsilon_{\mathrm{F}} = -\mu_{\mathrm{h}}$) and hole concentration. In accordance with Eqs. (13) and (28), the shift in the Fermi level $\Delta\varepsilon_{\mathrm{F}}^{\mathrm{int}}$ is given by

$$\Delta\varepsilon_{\mathrm{F}}^{\mathrm{int}} = -0.5\Delta\mu_{\mathrm{V}} = -0.5kT \ln \gamma_{\mathrm{V}}. \qquad (35)$$

When correlation effects are small, $\Delta\varepsilon_{\mathrm{F}}^{\mathrm{int}}$ can be approximated by

$$\Delta\varepsilon_{\mathrm{F}}^{\mathrm{int}} = 0.5kT \ln\left(p_{\mathrm{f}} + p_{\mathrm{R}} e^{\frac{\Delta E_{\mathrm{VR}}}{kT}} + p_{\mathrm{2R}} e^{\frac{\Delta E_{\mathrm{V2R}}}{kT}}\right). \qquad (36)$$

Earlier, we derived this equation for the special case of $\Delta E_{\mathrm{VR}} = \Delta E_{\mathrm{V2R}}$ and low $x$ [18].

Furthermore, the hole concentration can be expressed through $\Delta\varepsilon_{\mathrm{F}}^{\mathrm{int}}$ as follows

$$c_{\mathrm{h}} = c_{\mathrm{h}}^{0} \exp\left(-\frac{\Delta\varepsilon_{\mathrm{F}}^{\mathrm{int}}}{kT}\right) = c_{\mathrm{h}}^{0}\sqrt{\gamma_{\mathrm{V}}}, \qquad (37)$$

where $c_{\mathrm{h}}^{0}$ is the hole concentration in the absence of inter-defect interactions.

Eq. (36) indicates that trapping shifts the Fermi level to higher energies in the band gap (as predicted in Ref. [18]), thereby reducing the hole concentration (see Eq. (37)). The effect of trapping on $\varepsilon_{\mathrm{F}}$ and $c_{\mathrm{h}}$ diminishes with increasing temperature. For example, at $x = 0.2$ and realistic energy parameters ($\Delta E_{\mathrm{VR}} = 0.4$ eV, $\Delta E_{\mathrm{V2R}}/\Delta E_{\mathrm{VR}} = 1.5$), $\Delta\varepsilon_{\mathrm{F}}^{\mathrm{int}}$ decreases from 0.20 to 0.16 eV and $c_{\mathrm{h}}/c_{\mathrm{h}}^{0}$ increases from $10^{-1.7}$ to $10^{-0.7}$ as temperature rises from 600 to 1200 K.

It should be noted that the effect of inter-defect interactions and impurity disorder on oxidation is essential for understanding the behavior of hole conduction. In particular, the oxidation enthalpy $\Delta H_{\mathrm{ox}}$ provides a direct contribution to the effective activation energy of hole conductivity $\sigma_{\mathrm{h}}$. This contribution, $\Delta E_{\mathrm{a}}^{\sigma\mathrm{h}}$, resulting from the $c_{\mathrm{h}}(T)$ dependence, is

$$\Delta E_{\mathrm{a}}^{\sigma\mathrm{h}} = \frac{\Delta H_{\mathrm{ox}}^{0} + \Delta H_{\mathrm{ox}}^{\mathrm{int}}}{2}, \qquad (38)$$

where $\Delta H_{\mathrm{ox}}^{0}$ is the term associated with the equilibrium constant $K_{\mathrm{ox}}^{0}$, independent of inter-defect interactions.

As discussed above, at realistic trapping energies, $\Delta H_{\mathrm{ox}}^{\mathrm{int}}$ can be significant and strongly depend on dopant content (see Fig. 11b), resulting in the corresponding behavior of hole conductivity and its activation energy.

Fig. 12 shows that the dependence of the hole concentration on $x$ is strongly affected by the trapping energy ratio $\Delta E_{\mathrm{V2R}}/\Delta E_{\mathrm{VR}}$. For $\Delta E_{\mathrm{V2R}}/\Delta E_{\mathrm{VR}} \approx 1$, $c_{\mathrm{h}}$ increases monotonically with increasing $x$; for larger ratios, a maximum appears on the dependence $c_{\mathrm{h}}(x)$ at low $x$. Note that the monotonic behavior of $c_{\mathrm{h}}(x)$ at not too high $\Delta E_{\mathrm{V2R}}/\Delta E_{\mathrm{VR}}$ ratios correlates with the observed behavior of the hole conductivity $\sigma_{\mathrm{h}}(x)$ in Y- and In-doped $BaSnO_3$ [35,36]. A detailed analysis of $\sigma_{\mathrm{h}}$ in these oxides requires elucidating the mechanisms of hole mobility and the effects of impurity disorder on $\sigma_{\mathrm{h}}$, which is beyond the scope of this work. However, at not too large energy barriers for hole polaron hopping, the dependence $\sigma_{\mathrm{h}}(x)$ is largely determined by the $c_{\mathrm{h}}(x)$ behavior. This can be easily understood by analogy with the behavior of proton hopping conductivity in perovskites under consideration [18,19]. Using the effective media approximation, we demonstrated that at low hopping barriers, the carrier' mobility weakly depends on $x$ outside the region of low $x$ [18,19]. Since the estimated barriers for hole polaron hopping in acceptor-doped $BaSnO_3$ are comparable to $kT$ (~ 0.1 eV) [45], we can expect that the dependences $\sigma_{\mathrm{h}}(x)$ and $c_{\mathrm{h}}(x)$ behave similarly at relevant temperatures.

Concluding this section, we would like to note that the obtained results on oxidation, predicting the position of $\varepsilon_{\mathrm{F}}$ and $c_{\mathrm{h}}$ behavior, can be important for understanding the performance of electrochemical devices based on the studied oxides (see, e.g., Refs. [1,18,48]).

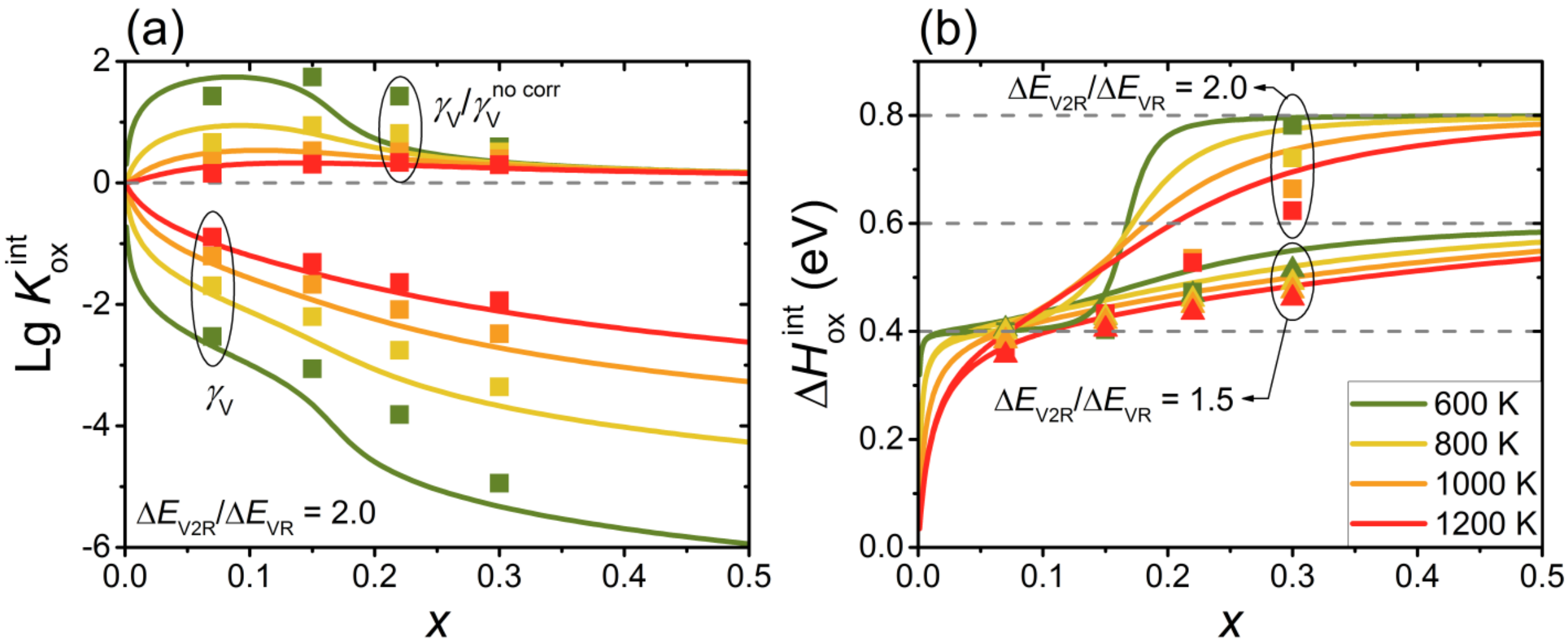

**Fig. 11.** Thermodynamic parameters of the oxidation reaction related to inter-defect interactions. (a) Component of the equilibrium constant $K_{\mathrm{ox}}^{\mathrm{int}} = \gamma_{\mathrm{V}}$ and (b) oxidation enthalpy $\Delta H_{\mathrm{ox}}^{\mathrm{int}}$ as a function of dopant content $x$ ($\Delta E_{\mathrm{VR}} = 0.4$ eV, $\Delta E_{\mathrm{V2R}}/\Delta E_{\mathrm{VR}} = 1.5$ and 2). The ratio of the activity coefficients $\gamma_{\mathrm{V}}/\gamma_{\mathrm{V}}^{\mathrm{no\ corr}}$, shown in (a), illustrates the effect of Fermi-type correlations on the equilibrium constant ($\gamma_{\mathrm{V}}$ and $\gamma_{\mathrm{V}}^{\mathrm{no\ corr}}$ are calculated with and without correlations). Symbols represent the parameters calculated with the non-uniform dopant distribution, corresponding

to a frozen high-temperature configuration (see Section 3.4). Dashed horizontal lines in (a) and (b) correspond to $K_{\text{ox}}^{\text{int}} = \gamma_{\text{V}} = 1$ and the values of the trapping energies ($\Delta E_{\text{VR}}$ and $\Delta E_{\text{V2R}}$), respectively.

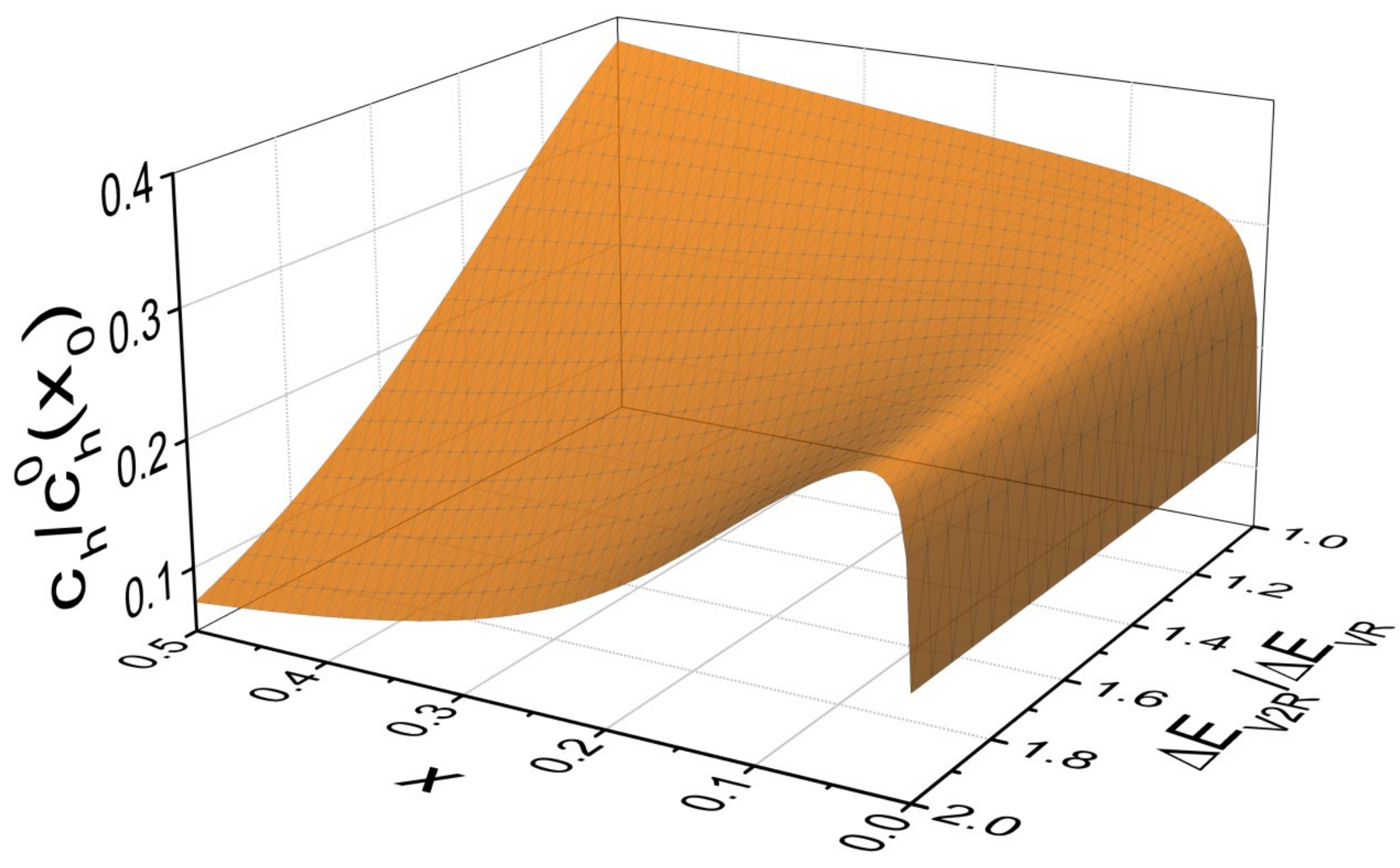

**Fig. 12.** Dependence of the normalized hole concentration $c_{\text{h}}/c_{\text{h}}^{0}(x_0)$ on dopant content $x$ and $\Delta E_{\text{V2R}}/\Delta E_{\text{VR}}$ ratio ($T = 1000$ K, $\Delta E_{\text{VR}} = 0.4$ eV). $c_{\text{h}}^{0}(x_0)$ is the hole concentration at $x_0 = 0.05$ in the absence of inter-defect interactions.

### 3.6 Comparison with experiment and other simulations

Our results on the distribution of oxygen vacancies can be compared with the findings of the NMR studies of local structure in acceptor-doped perovskites. Oikawa et al. [28] studied the $^{45}$Sc NMR spectra of vacuum-dried $BaZr_{1-x}Sc_xO_{3-\delta}$ ($x = 0.01$–$0.15$) and determined the integral intensities of the peaks assigned to the five- and six-oxygen-coordinated Sc. Given the relatively low dopant content in the studied samples [28], we will use for comparison our results for the case of weak correlations. For the uniform distribution of impurities, the fractions of the $RO_6$ and $RO_5$ groups are given, respectively, by

$$[x(1-\langle n_{\text{V2R}}\rangle) + (1-x)(1-\langle n_{\text{VR}}\rangle)]^6, \tag{39}$$

$$6[x\langle n_{\text{V2R}}\rangle + (1-x)\langle n_{\text{VR}}\rangle][x(1-\langle n_{\text{V2R}}\rangle) + (1-x)(1-\langle n_{\text{VR}}\rangle)]^5. \tag{40}$$

Fitting the experimental data [28] with Eqs. (39) and (40) yields $\Delta E_{\text{VR}}/kT \approx 3.0$ for $\Delta E_{\text{V2R}}/\Delta E_{\text{VR}} \approx 1.5$ and $\Delta E_{\text{VR}}/kT \approx 2.6$ for $\Delta E_{\text{V2R}}/\Delta E_{\text{VR}} \approx 2.0$. If we assume that the defect configuration is frozen at 1473 K (at which the samples were vacuum dried for 22 h [28]), then $\Delta E_{\text{VR}} \approx 0.3$-$0.4$ eV. It should be noted that variations in the parameter $\Delta E_{\text{V2R}}/\Delta E_{\text{VR}}$, in contrast to $\Delta E_{\text{VR}}/kT$, have little effect on the accuracy of the fitting. The determined trapping energy $\Delta E_{\text{VR}}$ is close to the values obtained by ab initio calculations for Sc-doped $BaZrO_3$ (0.37 eV [12] and 0.29 eV [14]). For Y-doped $BaZrO_3$, Draber et al. [24] obtained the $\Delta E_{\text{V2R}}/\Delta E_{\text{VR}}$ ratio of ~ 1.5.

Fig. 13a shows very good agreement between our theoretical results and the experimental data [28] for both values of the ratio $\Delta E_{V2R}/\Delta E_{VR}$ (1.5 and 2.0). As Sc substitution increases, the fraction of $RO_5$ groups increases due to the preferable location of oxygen vacancies near Sc ions. The ratio of the 5 to 6 oxygen-coordinated Sc increases with increasing $x$, significantly exceeding the ratio at uniform distribution of vacancies (Fig. 13b). The latter indicates a substantial effect of vacancy trapping on their distribution. A similar qualitative trend in the behavior of the 5:6 coordinated Sc ratio as a function of $x$ was reported by Buannic et al. [29] for Sc-doped $BaZrO_3$.

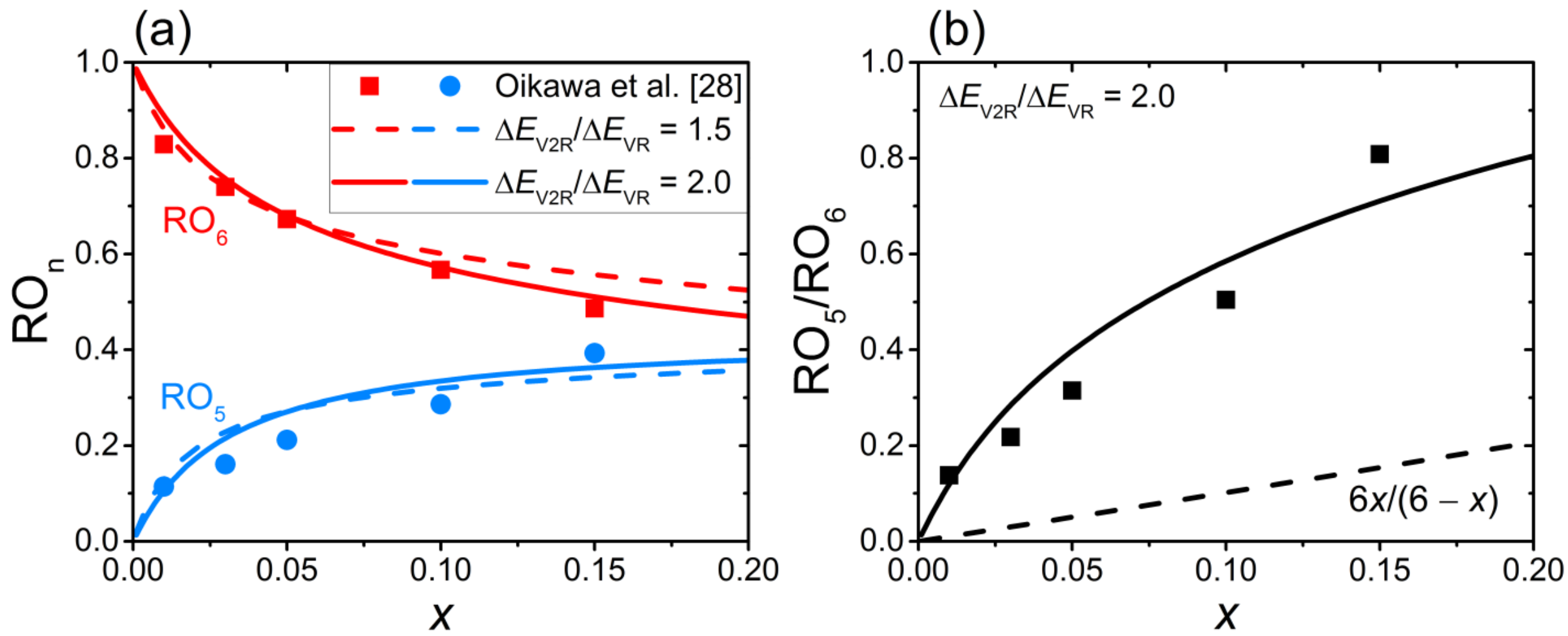

**Fig. 13.** (a) Fractions of $RO_5$ and $RO_6$ groups and (b) their ratio as a function of dopant content $x$. Lines represent theoretical dependencies for two $\Delta E_{V2R}/\Delta E_{VR}$ ratios: 1.5 ($\Delta E_{VR}/kT = 3.0$) and 2.0 ($\Delta E_{VR}/kT = 2.6$). The trapping energies ($\Delta E_{VR}/kT$) were determined from fitting of the experimental data [28]. Symbols correspond to the experimental values of five- and six-coordinated scandium ($ScO_5$ and $ScO_6$) in $BaZr_{1-x}Sc_xO_{3-\delta}$ extracted from the $^{45}Sc$ NMR spectra [28]. The dashed black line in (b) indicates the ratio for the uniform distribution of vacancies over oxygen sites ($6x/(6-x)$).

Buannic et al. [29] also analyzed the $^{17}O$ NMR spectra of $BaZr_{1-x}Sc_xO_{3-\delta}$, assigning relative peak intensities to the Zr-O-Zr, Zr-O-Sc and Sc-O-Sc environments, which roughly correspond to the probabilities $p_f$, $p_R$ and $p_{2R}$, respectively. For samples with 15 % Sc, the observed probabilities agree with a uniform Sc distribution, but at 30 % Sc, the experimental probability $p_{2R}$ is lower, indicating a reduced concentration of R-O-R groups. The latter correlates with the high-temperature dopant distribution simulated by Kasamatsu et al. [30] (see Section 3.4). It should be noted, however, that the experimental probabilities [29] $p_f$, $p_R$ and $p_{2R}$ do not satisfy the condition $2x = p_R + 2p_{2R}$ at $x = 0.3$ and therefore cannot be considered as reliable descriptors of the dopant distribution.

We further compare our findings obtained within our model with the combined DFT and Monte Carlo simulations [31]. Hoshino et al. [31] calculated the equilibrium defect configurations in 22.2% Sc-doped $BaZrO_3$ over a wide temperature range. The authors reported an almost uniform impurity distribution at ~ 1900 K. Fig. 14 presents the temperature dependencies of the oxygen vacancy concentrations computed in Ref. [31] for a frozen high-temperature dopant configuration. We calculated the same dependencies within our model analytically taking into account weak inter-vacancy correlations and by Monte-Carlo simulations considering both weak and strong correlations. The results are presented for the energy parameters $\Delta E_{VR} = 0.3$ eV and $\Delta E_{V2R}/\Delta E_{VR} = 2.0$, which were chosen based on our fitting of the experimental data (see Fig. 13) as well as the DFT calculations for Sc-doped $BaZrO_3$ [12,14]. Fig. 14 shows that the vacancy concentrations calculated for the case of strong inter-vacancy correlations almost perfectly match those reported in Ref. [31], while for weak correlations the results are noticeably different. Specifically, strong correlations result in a decrease in the concentration of R-V-R complexes and an increase in the concentration of R-V-B complexes at low *T*. This indicates the importance of inter-site inter-vacancy repulsion effects in heavily acceptor-doped oxides, especially at low temperatures (see Section 3.2). It should be emphasized, however, that the calculated equilibrium defect distribution is unlikely to be achieved in experiments at low temperatures, when the diffusivity of oxygen vacancies is low.

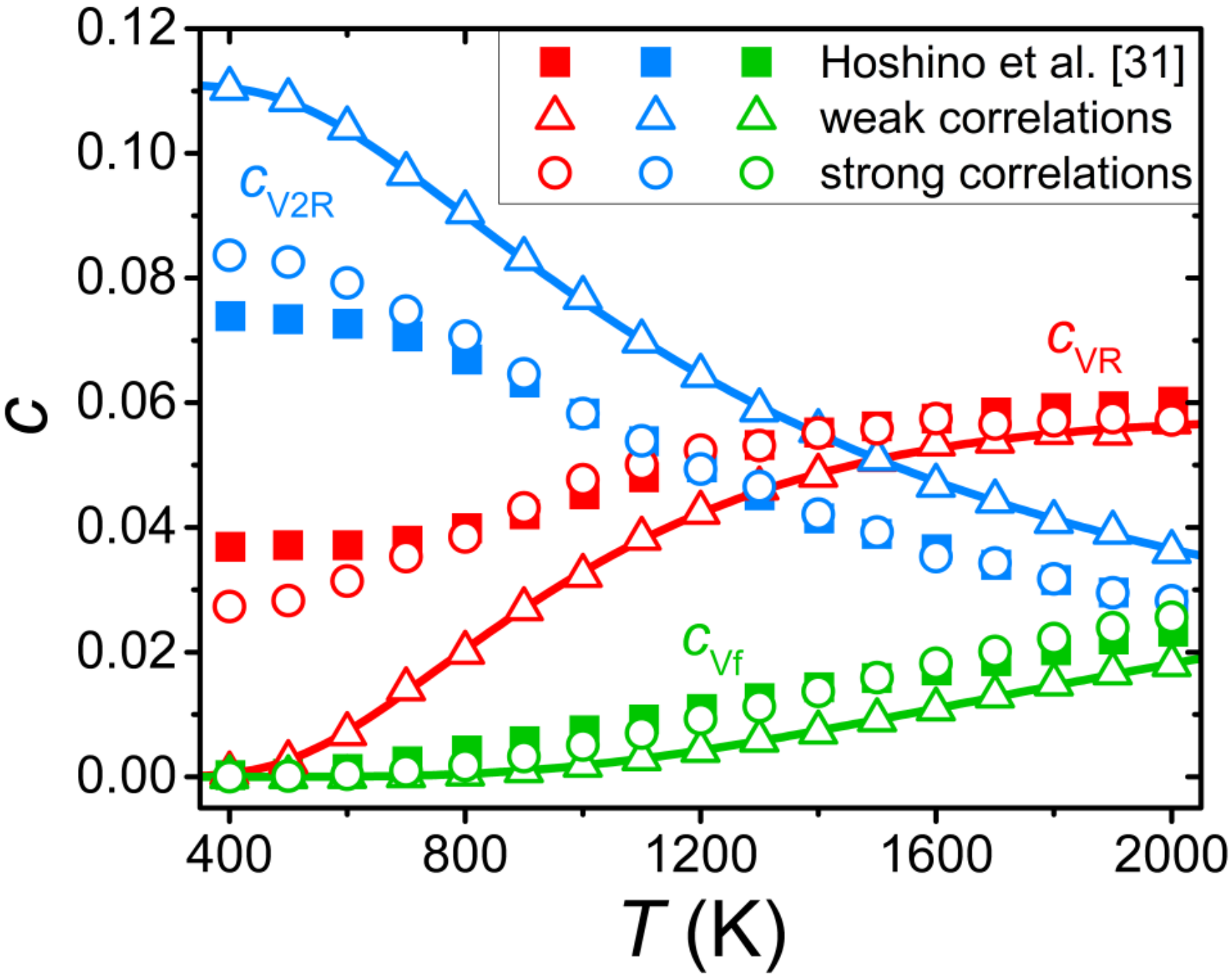

**Fig. 14.** Concentrations of acceptor-bound ($c_{VR}$ and $c_{V2R}$) and free ($c_{Vf}$) oxygen vacancies in 22.2% Sc-doped $BaZrO_3$ as a function of temperature *T*. Closed squares correspond to the results of the combined DFT and Monte Carlo simulations [31]. Solid lines and open symbols represent the results of analytical theory and Monte-Carlo modeling, respectively, obtained within our model at $\Delta E_{VR} = 0.3$ eV and $\Delta E_{V2R}/\Delta E_{VR} = 2.0$. Analytical calculations account for weak inter-vacancy correlations, and Monte-Carlo simulations incorporate both weak and strong correlations.

### 3.7 A brief discussion of the proposed model: limitations and possibilities

Our considerations are based on the model proposed to explore the effects of oxygen vacancies interaction with frozen-in acceptor impurities and with each other (see Section 2.1). In this model, the interaction with impurities and defect-induced lattice distortions is described using a three-level energy spectrum. These levels correspond to the three possible sets of the B-sublattice cations in the nearest environment of vacancies. The concentration and distribution of acceptor impurities are incorporated into defect thermodynamics through the probability of finding these sets. At first glance, such an approach seems oversimplified, since the model energy spectrum of vacancies does not depend on dopant content and configuration of impurities beyond the first neighbors.

However, similar models, treating acceptor-bound states of protons and vacancies even in a simpler form, provided a good description of hydration and proton transport in a number of acceptor-doped perovskites [10,17-19]. This indicates that the contribution of long-range inter-defect interactions is not critical for the defect thermodynamics of the studied compounds. The pivotal role of the interaction between vacancies and their local environment is largely related to the high static permittivity $\varepsilon_0$ of the considered wide-gap perovskites, which reduces the energy of long-range Coulomb interactions [17].

Another model limitation is related to long-range deformation interactions. These interactions can substantially contribute to the defects energy at a considerable ionic radii $R_i$ mismatch between dopants and substituted host cations. According to the DFT results [24,49], in perovskites with large-$R_i$ dopants, the trapping energies of vacancies on the dopant nearest- and second nearest-neighbor oxygen sites can be comparable. In this case, our model is directly inapplicable, since the trapping zones for vacancies overlap even at moderate dopant content $x$, and the energy spectrum of defect states becomes more complex and $x$-dependent. Consideration of such a problem would require compound-specific analysis with a low degree of generality, which is not in line with the goals of this work. It should be also noted that dopants with large $R_i$ typically exhibit low solubility and tend to substitute A-sublattice cations in the studied perovskites [50-52], which makes such materials less suitable for clean energy applications.

Obviously, the exploited approach cannot describe long-range effects, such, e.g., as ordering of oxygen vacancies. Note, however, that even in the case of materials with orderable defects, models with several defect energy states can be useful for analyzing bulk properties beyond phase transition regions. For example, a two-level model for orderable oxygen ions in the basal plane of $YBa_2Cu_3O_{7-\delta}$ allowed us to describe $p_{O2}$-$T$-$\delta$ diagrams and oxidation of this high-$T_c$ oxide over a wide range of external conditions (with worse than the mean-field and Bethe-Kikuchi approximations, but acceptable accuracy), see [53] and Refs therein.

The inter-vacancy interaction in the proposed model is treated as infinite short-range repulsion: (1) on-site (Fermi-type correlations) and (2) on nearest neighboring sites. The latter reflects the high Coulomb interaction energy and expectable instability of an oxygen octahedron with two such vacancies in a cubic perovskite. Our results show that both types of inter-vacancy interaction can significantly affect defect thermodynamics, local structure, and oxidation. Fermi-type correlations are most pronounced at relatively low dopant content $x$ within a narrow $x$ range (see Sections 3.2, 3.5). The effect of inter-site inter-vacancy repulsion increases with increasing $x$ and can become noticeable even at moderate doping levels in the low-temperature region (see Sections 3.2, 3.6).

Despite the above-mentioned limitations, the proposed model is useful not only for qualitative, but, in some cases, for quantitative analysis of oxide properties (at moderate dopant content). The considerations based on this model have allowed us to elucidate the influence of vacancy-impurity and inter-vacancy interactions on the local structure and oxidation of $AB_{1-x}R_xO_{3-\delta}$ perovskites. The obtained results describe well the NMR experimental data on local structure and correlate with the hole conductivity data for acceptor-doped perovskites (see Section 3.6). We also note that our predictions for the oxygen-vacancy distribution in 22.2% Sc-doped $BaZrO_3$ are consistent with the results of computationally intensive combined DFT and Monte-Carlo simulations (Section 3.6). Additionally, we have revealed the influence of the frozen non-uniform dopant distribution on the studied properties.

Furthermore, the model has predicted several interesting physical effects. In particular, we have shown the possibility of significant on-site Fermi-type correlations at relatively low dopant content and elucidate their implications. One of the most important predictions is related to the oxidation of acceptor-doped perovskites: we have revealed the possibility of a strong and unusual dependence of oxidation enthalpy and hole concentration on dopant content.

Finally, it should be noted that the proposed model can also be useful for a fairly wide class of problems in the theory of defect and transport phenomena in acceptor-doped perovskites. In particular, as we mentioned in Section 3.5, the results on oxidation are essential for understanding transport processes and functioning of the studied oxides in various electrochemical devices. Another potential application of the model − important for employing these oxides as proton-conducting materials − is the analysis of their properties (related, e.g., to hydration) in hydrogen-containing atmospheres. Hydration theory taking into account the effects of acceptor-bound states and briefly considering the role of 3-particle defect complexes (including R-V-R) was previously developed in our work [17]. Several important implications of 3-particle complexes were revealed [17], and further developments of the hydration theory using the new findings presented here is under way. We would also like to point out that the proposed approach can be instrumental in developing a theory of transport phenomena of oxide-ion charge carriers. Sketches of such a theory for the oxygen-vacancy conductivity, mobility and oxide-ion thermopower can be found in our recent papers [54,55].

## 4 Concluding remarks

In summary, we have investigated the effects of inter-defect interaction on the properties of wide-gap $AB_{1-x}R_xO_{3-\delta}$ perovskites under dry oxidizing conditions.

Using the developed statistical theory and Monte Carlo simulations, we have elucidated the influence of these interactions, along with quenched impurity disorder, on the oxygen-vacancy distribution, local coordination of host ions and defects, and oxidation.

The effects of vacancy-impurity and vacancy-vacancy interactions manifest themselves differently depending on external conditions and dopant content $x$. (a) Interaction between vacancies and impurities generally exerts a greater influence on the studied properties compared to inter-vacancy interaction. (b) The effect of on-site Fermi-type correlations, arising at high vacancy occupancy of oxygen sites surrounded by two impurities, can be considerable within a narrow dopant range at relatively low $x$. (c) The impact of short-range inter-site inter-vacancy correlations increases with $x$ and can become noticeable even at moderate doping levels in the low-temperature region.

The main results regarding the studied properties can be summarized as follows:

1. The concentrations of vacancies in various cation environments behave differently depending on dopant content. The fraction of vacancies located near a single impurity varies non-monotonically with $x$, reaching a maximum at relatively low $x$. The fraction of vacancies surrounded by two impurities increases with $x$, and, under typical energy parameters, these vacancies become dominant even at moderate $x$ values;
2. Inter-defect interactions result in the preferential formation of oxygen vacancies in the vicinity of acceptor impurities. Stronger vacancy-impurity interaction renders the local coordination increasingly non-uniform, while inter-vacancy interaction has the opposite and a significantly less pronounced effect;
3. Inter-vacancy correlations result in the dependence of short-range order parameters on dopant content. At low doping levels, the effect can be largely due to Fermi-type correlations, whereas at high dopant content, inter-site vacancy repulsion plays a dominant role;
4. Non-uniform dopant allocation, which can result from a sample preparation method, can noticeably influence the vacancy distribution over different types of oxygen sites, but has little effect on the local coordination and oxidation;
5. Inter-defect interactions considerably affect the oxidation behavior of acceptor-doped perovskites, reducing the oxidation constant. These interactions can result in a strong and unusual dependence of oxidation enthalpy and hole concentration on dopant content.

The obtained results elucidate the implications of defect interactions and impurity disorder in acceptor-doped oxides, enabling the optimization of their properties required for applications through the selection of dopant type and content.

## Conflicts of interest

There are no conflicts of interest to declare.